\documentclass[%
 reprint,
 amsmath,amssymb,
 aps,
 pra,
]{revtex4-1}

\usepackage{dcolumn}%
\usepackage{bm}%

\usepackage{amsmath,amsthm}
\usepackage{amssymb}
\usepackage{amscd}
\usepackage{wasysym}
\usepackage[utf8]{inputenc}
\usepackage[T1]{fontenc}
\usepackage{ae,aecompl}
\usepackage{hyperref}
\usepackage{sidecap}
\usepackage{color}

\usepackage[pdftex]{graphicx}%
\DeclareGraphicsExtensions{.pdf}

\usepackage[pdftex,usenames,dvipsnames]{xcolor}
\usepackage{xargs}
\usepackage[colorinlistoftodos,prependcaption,textsize=small]{todonotes}
\newcommandx{\todog}[2][1=]{\todo[linecolor=Green,backgroundcolor=Green!25,bordercolor=Green,#1]{#2}}
\newcommandx{\todop}[2][1=]{\todo[linecolor=Magenta,backgroundcolor=Magenta!50,bordercolor=Magenta,#1]{#2}}

\newcommand{\abs}[1]{\vert#1\vert}
\newcommand{\be}{\begin{equation}}
\newcommand{\ee}{\end{equation}}
\newcommand{\bee}{\begin{equation*}}
\newcommand{\eee}{\end{equation*}}
\newcommand{\bea}{\begin{eqnarray}}
\newcommand{\eea}{\end{eqnarray}}

\renewcommand{\vec}[1]{\boldsymbol{#1}}

\newcommand{\LL}{\mathcal{L}}

\newtheorem{theorem}{Theorem}

\begin{document}

\title{Collective effects of link failures in linear flow networks}%

\author{Franz Kaiser}
\thanks{FK and JS contributed equally to this work.}
\affiliation{Forschungszentrum J\"ulich, Institute for Energy and Climate Research -
	Systems Analysis and Technology Evaluation (IEK-STE),  52428 J\"ulich, Germany}
\affiliation{Institute for Theoretical Physics, University of Cologne,
		50937 K\"oln, Germany}
		
\author{Julius Strake}
\thanks{FK and JS contributed equally to this work.}
\affiliation{Forschungszentrum J\"ulich, Institute for Energy and Climate Research -
	Systems Analysis and Technology Evaluation (IEK-STE),  52428 J\"ulich, Germany}
\affiliation{Institute for Theoretical Physics, University of Cologne,
		50937 K\"oln, Germany}
		
\author{Dirk Witthaut}
\affiliation{Forschungszentrum J\"ulich, Institute for Energy and Climate Research -
	Systems Analysis and Technology Evaluation (IEK-STE),  52428 J\"ulich, Germany}
\affiliation{Institute for Theoretical Physics, University of Cologne,
		50937 K\"oln, Germany}

\date{\today}

\begin{abstract}
The smooth operation of supply networks is crucial for the proper functioning of many systems, ranging from biological organisms such as the human blood transport system or plant leaves to man-made systems such as power grids or gas pipelines. Whereas the failure of single transmission elements has been analysed thoroughly for power grids, the understanding of multiple failures is becoming more and more important to prevent large scale outages with an increasing penetration of renewable energy sources. In this publication, we examine the collective nature of the simultaneous failure of several transmission elements. In particular, we focus on the difference between single transmission element failures and the collective failure of several elements. We demonstrate that already for two concurrent failures, the simultaneous outage can lead to an inversion of the direction of flow as compared to the two individual failures and find situations where additional outages may be beneficial for the overall system. In addition to that, we introduce a quantifier that performs very well in predicting if two outages act strongly collectively or may be treated as individual failures mathematically. Finally, we extend on recent progress made on the understanding of single link failures demonstrating that multiple link failures may be treated as superpositions of multiple electrical dipoles for lattice-like networks with collective effects completely vanishing in the continuum limit. Our results demonstrate that the simultaneous failure of multiple lines may lead to unexpected effects that cannot be easily described using the theoretical framework for single link failures.
\end{abstract}

\maketitle

\section{Introduction}
\label{sec:intro}

The failure of links can impede the operation of supply networks leading to potentially critical events; cascading failures in power grids can cause power outages affecting millions of households~\cite{atputharajah_power_2009,Pour06} and embolism in humans and plants may result in strokes~\cite{shih_smallest_2013,Schaffer2006} or leaf death~\cite{brodribb_revealing_2016}. Such events are typically caused by the failure of one or few transportation links~\cite{Pour06}. 

To prevent power outages caused by single link failures, power transmission grid operators typically run the grids $N-1$ secure which means that a single failing transmission or generation element does not prevent stable operation of the power grid~\cite{Wood14}. However, an increased risk of extreme weather events caused by climate change raises the risk several transmission elements failing thus leading to power outages~\cite{campbell_weather-related_nodate}. In addition to that, future power systems with a high share of renewable energy sources will have to transport power over long distances using long transmission lines, thus also increasing the risk of dependent link failures. For these reasons, grid operators are now encouraged to take specific dangerous $N-2$ contingencies into account~\cite{operatorsN2outages2006} and an increase in correlation between transmission outages, e.g.\ through more extreme weather events, was recently associated with an increased risk of cascading outages, raising the relative contribution of multiple link failures to such cascades~\cite{clarfeld_risk_2019}.

Many computational approaches towards studying and classifying $N-2$ outages have been developed in order to study contingencies that result in additional overloads~\cite{turitsyn_fast_2012,kaplunovich_fast_2013,Davis2011,weckesser_identifying_2018}. Nevertheless, such outages still lack a fundamental theoretical understanding. Basic mathematical tools have been developed extending the concept of Line Outage Distribution Factors (LODFs) originally used for single link contingencies~\cite{Wood14} to include multiple link failures thereby allowing for a mathematical description of these contingencies~\cite{guler_generalized_2007,Guo09,soltan_quantifying_2016}. These tools demonstrate that the nature of multiple outages may be fundamentally different from the outage of single lines, thus making a direct transfer of understanding and intuition developed for single link failures~\cite{strake2018non} difficult. In particular, multiple outages can enhance or attenuate each other in a counterintuitive manner. Which topologies drive such phenomena and which ones prevent them from happening is at present not fully understood.

In this article we analyse the collective nature of $N - 2$ failures in linear flow networks, which describe different systems including AC power grids in the DC approximation~\cite{Wood14}. We demonstrate that two simultaneous failures can cause a disturbance which strongly differs from the sum of the disturbances induced by individual failures; they can amplify or attenuate the flow changes in a grid. In addition to that, we introduce a predictor which allows us to understand under which circumstances these collective effects play an important role and when they can be neglected. We then apply the predictor to different test grids and reveal its performance in forecasting collective effects for multiple link failures quantitatively, outperforming also distance measures proven to be good predictors in the case of single link failures. Finally, we extend on previous work~\cite{strake2018non} which successfully established an analogy between flow rerouting after single link failures and the fields of electromagnetic dipoles in regular grids, demonstrating that flows after multiple link failures may be treated as an overlay of multiple individual dipole fields in such grids in the continuum limit.

\section{link failures in linear flow networks}
\subsection{Fundamentals of linear flow networks}

Linear flow networks describe the operation of various types of systems including AC power grids \cite{Wood14,Purc05,Hert06}, DC electric circuits~\cite{bollobas1998,Dorfler2018,kirchhoff_ueber_1847}, hydraulic networks~\cite{Hwan96,diaz2016} and vascular networks of plants~\cite{Kati10}. In such networks, the flow $F_{m\rightarrow n}\in \mathbb{R}$ over a link $(m,n)$ is assumed to be linear in the potential or pressure drop along this link
\be
    F_{m \rightarrow n} = b_{mn} (\theta_m - \theta_n).
    \label{eq:flow-line}
\ee
In this article, we focus on applications to AC power grids, where $F_{m \rightarrow n}$ is the real power flow, $\theta_n\in\mathbb{R}$ is the voltage phase angle at node $n$ and $b_{mn}\in\mathbb{R}$ is proportional to the link's susceptance. We assume that the susceptance is independent of the direction of the link $b_{mn}=b_{nm}$ and that it vanishes if no link $(m,n)$ exists. In this context, the linear description is commonly referred to as the DC approximation due to its formal equivalence with DC resistor networks~\cite{Wood14,Purc05,Hert06}. This approximation is typically good for transmission grids with weak link loading, see Ref.~\cite{Purc05} for details. In hydraulic or vascular networks, $\theta_n$ denotes the pressure at node $n$ while the transmission capacity $b_{mn}$ depends on the geometry of a pipe or vein \cite{Hwan96,diaz2016,Kati10}. The flows are subject to the continuity equation which means that at each node of the grid the sum of the network flows must equal the inflow to the grid;
\be
    \sum_{n=1}^N F_{m \rightarrow n} = P_m.
    \label{eq:continuity}
\ee
The inflow $P_m$ is positive if a current, power or fluid is injected to the node and negative if it is withdrawn from the grid. 

Equations (\ref{eq:flow-line}) and (\ref{eq:continuity}) fully describe the state and the flow of the network once the link parameters $b_{mn}$ and the injections $P_m$ are given up to a constant phase shift applied to all voltage phase angles. We introduce a compact vectorial notation, summarising the nodal potentials or voltage phase angles in the vector $\vec \theta = (\theta_1,\ldots,\theta_N)^\top \in \mathbb{R}^N$ and the nodal injections in the vector $\vec P = (P_1,\ldots,P_N)^\top \in \mathbb{R}^N$. Here and in the following sections, the superscript `$\top$' denotes the transpose of a vector or matrix. We further label all lines in the grid by $l = 1,\ldots,M$ and fix an orientation for each link. Summarising all link flows in the vector $\vec F = (F_1,\ldots,F_M)^\top \in \mathbb{R}^M$, Equation~(\ref{eq:flow-line}) reads as 
\be
    \vec F = \vec B_d \vec I^\top \vec \theta,\nonumber
\ee
with the diagonal matrix of link strengths $\vec B_d = {\rm diag}(b_1,b_2,\ldots,b_M) \in \mathbb{R}^{M \times M}$. Furthermore, we made use of the node-edge incidence matrix $\vec I \in \mathbb{R}^{N \times M}$ commonly used in graph theory. It establishes a correspondence between the nodes in the graph and the edges connecting them and has the components \cite{Newm10}
\be
   I_{n,\ell} = \left\{
   \begin{array}{r l}
      1 & \; \mbox{if link $\ell$ starts at node $n$},  \\
      - 1 & \; \mbox{if link $\ell$ ends at node $n$},  \\
      0     & \; \mbox{otherwise}.
  \end{array} \right.
  \nonumber%
\ee
In the following, we use this matrix to assign an (arbitrary) orientation to each link in the network such that $F_{m\rightarrow n}=-F_{n\rightarrow m}$. Using the node-edge incidence matrix we can further rewrite the continuity equation (\ref{eq:continuity}) in the compact form
\begin{equation}
  \vec P = \vec I \vec F = \vec I \vec B_d \vec I^\top \vec \theta = \vec B \vec \theta.
  \label{eq:DCapprox}
\end{equation}
The matrix $\vec B = \vec I \vec B_d \vec I^\top \in \mathbb{R}^{N \times N}$ is commonly referred to as the nodal susceptance matrix in power engineering. Mathematically, $\vec B$ is a weighted Laplacian matrix~\cite{Newm10,merris1994} with components
\begin{equation}
  B_{m n} = \left\{
   \begin{array}{lll}
   \displaystyle\sum \nolimits_{s=1}^N b_{ns} &  \mbox{if } & m = n; \\ [2mm]
     - b_{mn} & & m \neq n.
   \end{array} \right. \label{eq:Bweighted}
\end{equation}
For a connected network, this matrix has one zero eigenvalue $\lambda_1=0$ with eigenvector $\vec v_1=\vec 1$ such that $\vec B \cdot\vec v_1=\vec 0$. For this reason, it is not invertible. However, the matrix inverse appears naturally in many different contexts involving the spreading of failures in networks. To be able to nevertheless study these processes, one typically considers the Moore-Penrose pseudoinverse $\bm{B}^\dagger$ which has properties similar to the actual inverse, see e.g. Ref.~\cite{Moore1920} for details. We are now ready to extend the notation to cover link failures as well, as covered in the next section.

\subsection{Single and double link failures}
\label{sec:single_and_double}
Assume that a single link $k$ in the network fails, thus loosing its ability to carry any flow. Since the network after the failure is still subject to the continuity equation~\eqref{eq:continuity}, the failure will cause the flows on other links to change to account for the remaining necessary transport. Assume that the new flows are given as $\hat{\vec F} = \vec F +\Delta \vec F$, where $\Delta \vec F$ is the vector of flow changes, and fulfil the continuity equation~\eqref{eq:DCapprox}
\bee
\bm P = \hat{\bm I}\hat{\vec F}
\eee
where $\hat{\bm I}$ is the node-edge incidence matrix of the network after removal of link $k$. In power engineering, the changes of flows are typically captured in a matrix of Line Outage Distribution Factors (LODFs) whose element $L_{l,k}$ describes the flow changes monitored on a link $l$ after another link $k$ fails. Suppose that $F^{(0)}_{k}$ is the flow on link $k$ before the outage. In general, we will use the superscript $^{(0)}$, i.e. round brackets, to indicate a flow before an outage. Then the LODF is defined by its elements~\cite{Wood14}
\begin{align}
    L_{l,k}:= \frac{\Delta F_{l}}{F^{(0)}_{k}}.
    \label{eq:LODF_PTDF}
\end{align}
Importantly, the LODF may also be expressed in purely algebraic form using the (pseudo) inverse of the graph Laplacian $\bm{B}$~\cite{Wood14}
\begin{align}
    L_{l,k}=b_{l}\frac{\bm d_{l}^\top\bm B^\dagger \bm d_{k}}{1-b_{k}\bm{d}_{k}^\top\bm B^\dagger \bm d_{k}}.
    \label{eq:LODF}
\end{align}
Here, we abbreviate the line susceptance $b_{l_1l_2}$ of a link $l=(l_1,l_2)$ by $b_{l}$. Furthermore, we defined a vector $\bm d_k \in \mathbb{Z}^{N}$ that characterises a link $k=(r,s)$ and has the entries $+1$ at position $r$, $-1$ at position $s$ and zero otherwise. Using the standard basis vectors in $\mathbb{R}^N$, this vector may be written as $\bm d_{k}=\bm{e}_r-\bm e_s$. In power engineering, the expression in the numerator is also referred to as the Power Transfer Distribution Factor (PTDF)~\cite{Wood14}. A PTDF between links $l$ and $k$ is calculated as
\begin{align*}
    \text{PTDF}_{l,k}=b_{l}\bm d_{l}^\top\bm B^\dagger \bm d_{k}
\end{align*}
and describes the flow changes on link $l$ upon a power transfer from the one end of link $k$ to the other one. PTDFs are typically defined for power injections and withdrawals at arbitrary nodes in the network~\cite{Wood14}. In the context of link failures, however, it is useful to restrict them to power injection and withdrawal taking place at the two ends of a link. For this reason, power injection vectors $\bm{d}_k$ have to correspond to the columns of the incidence matrix in our setup such that $\bm d_{k}=\bm I\cdot\bm e_{k}$, where $\bm e_k\in\mathbb{R}^{M}$ is the vector with entry one at position of edge $k=(r,s)$ and zero otherwise. 

On the other hand, the link failure may also be described on the nodal level. If we collect all changes in voltage phase angles after the failure of link $k$ in the vector $\vec \psi = \hat{\vec \theta} - \vec \theta$, again denoting phase angles after the failure by $\hat{\vec \theta}$, a Poisson-like equation describing the outage in terms of the phase differences may be derived~\cite{strake2018non},
\begin{align}
    \hat{\vec B}  \vec \psi = F_k^{(0)} \vec d_k.
    \label{eq:Poisson}
\end{align}
Here, $\hat{\vec{B}}$ is the Laplacian of the network after removal of link $k$ and $F_k^{(0)}$ is the pre-outage flow on link $k$. This equation was studied in the past in different settings~\cite{norman97,strake2018non}. The failure of single links is thus comparably well understood~\cite{strake2018non,Labavic2014,16redundancy}, whereas the simultaneous failure of multiple links was not yet studied to the same extend on a theoretical level.

For this reason, we now turn to the case of multiple link failures and derive an expression for the flow changes on the remaining lines in the grid. We will focus on the case of two outages for now. Naively, we could just superimpose the flow changes caused by the two individual outages as described by Equation~\eqref{eq:LODF_PTDF}. Assuming that two arbitrary lines $o$ and $k$ fail, this naive approach yields the following expression for the flow changes on link $l$
\begin{align}
    \Delta F_{l}^{\text{naive}} = L_{l,k} F^{(0)}_{k} + L_{l,o} F^{(0)}_{o}
     \label{eq:naive_flowchanges}
\end{align}
However, this approach neglects the effect of the outage of link $k$ on link $o$ and vice versa. To arrive at the correct formula, we need to consider this interaction as follows; the outage of link $o$ changes the flow on link $k$ by 
\begin{align*}
    \tilde{F}_{k} = F^{(0)}_{k} + L_{k,o} \tilde{F}_{o}.
\end{align*} 
An analogous expression holds for the effect of the outage of link $k$ on link $o$. Inserting these corrected flows into Equation~\eqref{eq:naive_flowchanges}, we arrive at the following result for the flow changes on link $l$~\cite{Davis2011}
\begin{align*}
    \Delta F_{l}=L_{l,k} \tilde{F}_{k}
     + L_{l,o} \tilde{F}_{o}.
\end{align*}
Finally, expanding this expression results in the following equation encoding the collective flow changes in a compact form
\begin{align}
    \Delta F_{l}&= 
    (L_{l,o},L_{l,k})
    &\begin{pmatrix}
    1&-L_{o,k}\\
    -L_{k,o}&1
    \end{pmatrix}^{-1}
    \begin{pmatrix}
    F^{(0)}_{o}\\F^{(0)}_{k}
    \end{pmatrix}. \label{eq:flow_changes_davis}
\end{align}
The resulting expression for $\Delta F_{l}$ is different from the simple linear combination \eqref{eq:naive_flowchanges} due to the interaction of the two failing lines $o$ and $k$, which is encoded in the matrix in the centre. More precisely, the collective effects are governed by the LODFs of the interacting lines $L_{o,k}$ and $L_{k,o}$ forming the off-diagonal elements of the matrix.
The inverse matrix in this formula may be calculated as
\begin{align*}
        &\begin{pmatrix}
    1&-L_{o,k}\\
    -L_{k,o}&1
    \end{pmatrix}^{-1}=\frac{1}{1-L_{o,k}L_{k,o}}    
    \begin{pmatrix}
    1&L_{o,k}\\
    L_{k,o}&1
    \end{pmatrix}.
\end{align*}
Thus, the flow changes on a link $l$ are given by
\begin{align*}
    \Delta F_{l} = & \LL(k,o) \cdot (L_{l,k}F^{(0)}_k+L_{l,o}F^{(0)}_o \\
    & \qquad \qquad + L_{l,k}L_{k,o}F_o^{(0)} +
    L_{l,o}L_{o,k}F_k^{(0)}),
\end{align*}
where $\LL(k,o)=\LL(o,k):=\frac{1}{1-L_{k,o}L_{o,k}}$ is a symmetric prefactor. 

The equation describing flow changes after the failure of two lines thus differs from a naive overlay of the two individual outages. In the following sections, we will demonstrate in which cases these collective effects resulting from the interaction of both outages are important and in which cases they may be neglected. 

\subsection{Elementary examples}
In this section, we elucidate different elementary examples that describe possible interactions between the individual outages and allow us to understand the role played by collective effects in more detail. 

\subsubsection{Amplifying single outages}

\begin{figure}[tb]
    \begin{center}
        \includegraphics[width=1.\columnwidth]{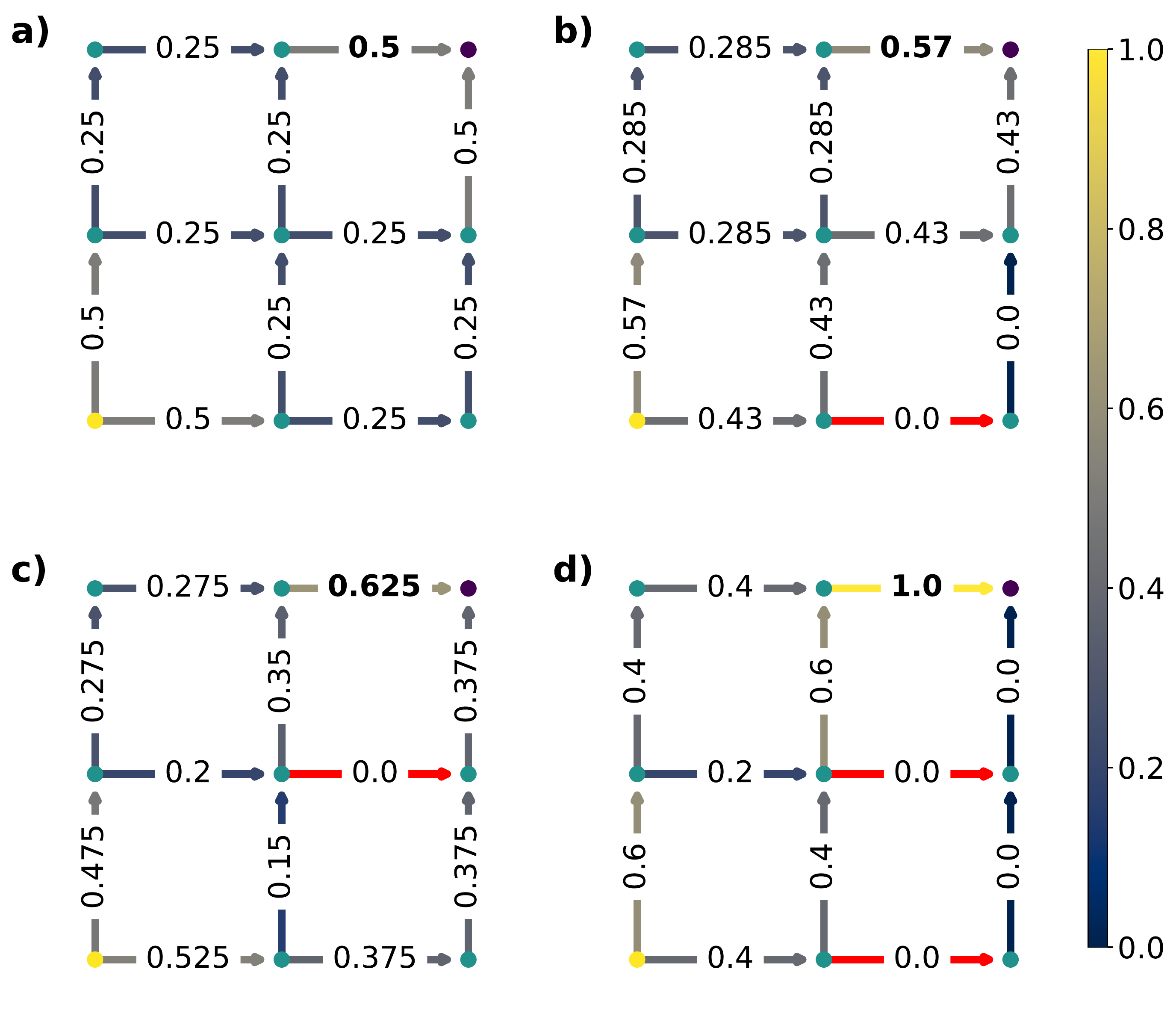}
    \end{center}
    \caption{
     Collective effects can amplify the flow changes after $N-2$ failures, thus increasing maximum link loading. The node in the lower left corner (yellow) is assumed to be a producer of one unit of power and the node in the upper right corner (purple) a consumer of the same amount. All links have a capacity of $b=1$ and the numbers on the edges indicate the (absolute) flow carried by the respective link with the arrow pointing in the direction of positive flows. In addition to that, the colour code ranging from black (no loading) to yellow (maximal loading) indicates the loading of the links. Bold face numbers indicate the link with the highest loading. The naive overlay may be calculated by subtracting from the flows observed in (b) and (c) the base indicated flows in a), thus calculating the flow changes, and then adding the result to a). This yields flows on the maximally loaded edge in d) of $\abs{F_{\text{max}}^{\text{naive}}}\approx 0.5+0.125+0.07= 0.72$. Thus, the naive overlay underestimates the flow on the maximally loaded edge in this case.
     }
    \label{fig:amplifying_outages}
\end{figure}

To start with, we present a case where the naive overlay of two individual outages underestimates the collective effects such that
\bee
\abs{\Delta F_l}\gg \abs{\Delta F_l^\text{naive}}
\eee
for some link $l$. An elementary example of a network where this is happening is shown in Fig.~\ref{fig:amplifying_outages} where the topology is given by a network consisting of $N=9$ nodes and $M=12$ links connecting them in a square grid. This initial setup is shown in panel a). Panels b) and c) show the flows flowing on each link (numbers on links) after the failure of two different individual links (coloured red). The bold number indicates the link with maximal flow for the given setup. Each single outage leads to a maximal flow on the top right link of $\abs{F_{\text{max}}} = 0.57$ and $\abs{F_{\text{max}}} = 0.625$ after the failure for panels b and c, respectively. Naively, we would thus expect the failure of both links to lead to a flow of 
\bee
\abs{F_{\text{max}}^{\text{naive}}} = \abs{\Delta F_{\text{max}}^{\text{naive}}} +\abs{F_{\text{max}}^{(0)}}\approx 0.125+0.07+0.5 = 0.72
\eee
by simply superposing the two individual outages. The actual outage of both links, however, results in a much higher flow on the link which reads as
\bee
\abs{F_{\text{max}}}=\abs{\Delta F_{\text{max}}}+ \abs{F_{\text{max}}^{(0)}}=1.0.
\eee
Whereas the two individual outages separately lead to a moderate increase in flow on the link with the maximal flow, their interaction results in a higher flow potentially reaching the link limit. If the flow on all links would be limited to, say, $F_{\text{limit}}=0.9$, the naive overlay would thus predict no overloading caused by the two link failures, whereas in fact, the link loaded maximally in panel d) breaks down in this case. This example demonstrates that the results predicted by the theory of single link failures may differ drastically from the correct calculation taking into account the collective effects. 

\subsubsection{Effect of individual outages exceeding simultaneous outage}

\begin{figure}[tb]
    \begin{center}
    \includegraphics[width=1.\columnwidth]{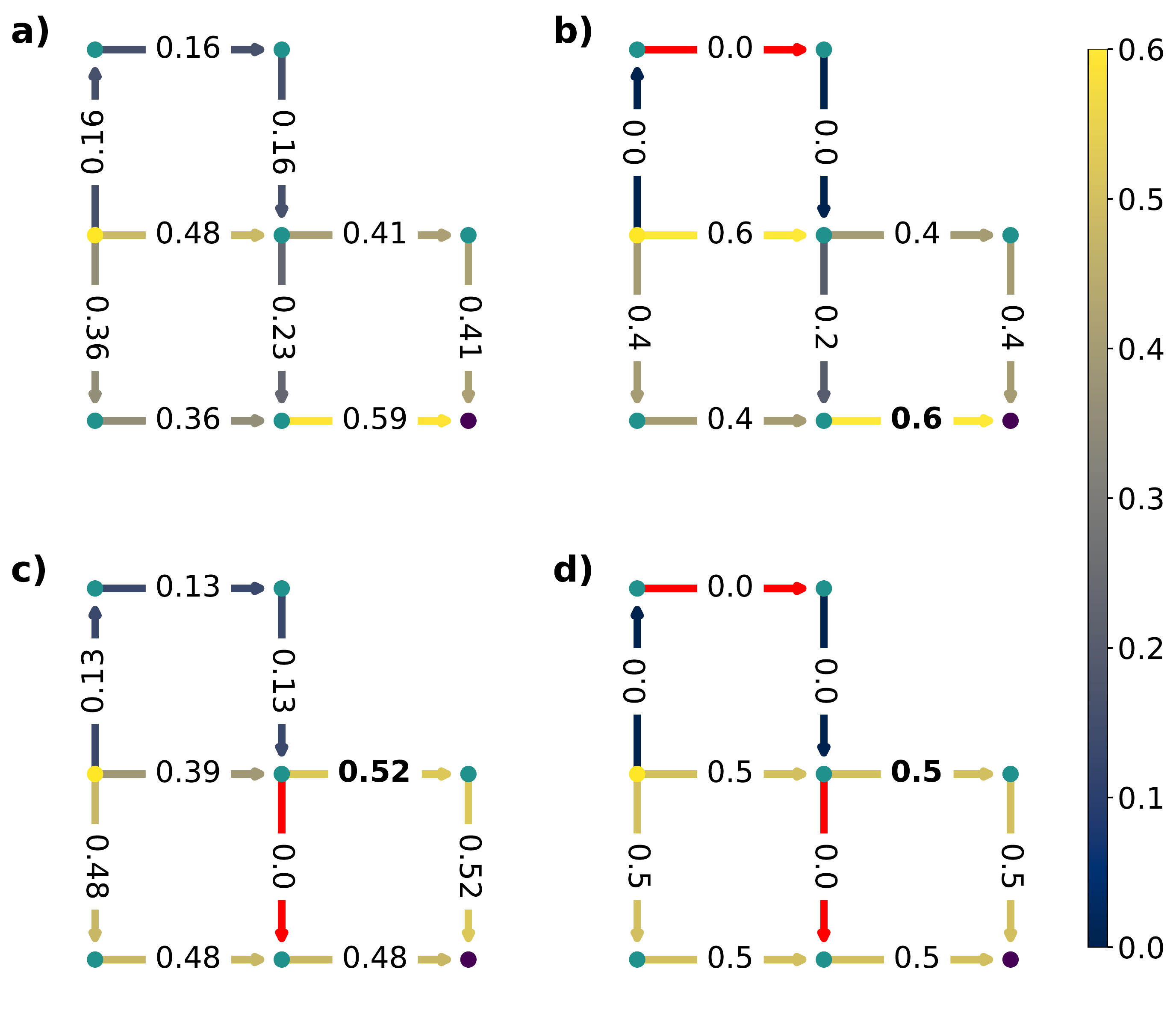}
    \end{center}
    \caption{
    Collective effects can attenuate flow changes after an $N-2$ failure, thus making the contingency less severe. Colour code of edges indicates the absolute flow on the link going from blue for no flow to yellow for links with maximum flow. The numbers on the links also represent the flow with the arrows pointing in the direction of positive flow. (a) Initial flow setup if there is a unit inflow at the yellow node on the left and a unit outflow at the purple node on the bottom right. (b) Flow setup after the failure of the top horizontal link (red). (c) Flow setup after the failure of the central, vertical link (red). (d) Flow setup after the failure of both links. While the edge with maximum absolute flow after the failure of the top link carries $\abs{F_\text{max,b}}=0.6$ units (b) and the maximum absolute flow after failure of the second link reads as $\abs{F_\text{max,c}}=0.52$ (c), this maximum after the failure of both links reads as $\abs{F_\text{max,d}}=0.5$. Thus, in both cases of individual failures, the failure of an additional link would be beneficial in terms of the maximal absolute flow in the network. This is a realisation of \textit{Braess's paradox}.
    }
    \label{fig:braess_realization}
\end{figure}

In addition to the effect presented in the last section, it may also happen that an additional outage is beneficial for the maximally loaded link. A minimal example is shown in Figure~\ref{fig:braess_realization}. In this setup, we have a unit inflow of power at the centre left node (coloured yellow) $P=+1$ and a unit outflow at the bottom right node (coloured purple) $P=-1$ whereas all other nodes neither consume nor create power $P=0$. The initial setup is shown in panel a), where again (absolute) flows are indicated as numbers on the edges as well as colour coded. Panels b) and c) show the network after the single outage of two different links (coloured red). The edge with the highest absolute flow is indicated by a bold face number in both cases. In panel b), we have $\abs{F_{\text{max,b}}}=0.6$ whereas for panel c) the maximum flow reads as $\abs{F_{\text{max,c}}}=0.52$. The situation after the simultaneous outage of both links is shown in panel d). The edge with maximum flow now carries an absolute flow of $\abs{F_{\text{max,d}}}=0.5$, i.e. the collective effects attenuate the flow on the edge with the highest flow compared to each individual outage. This effect can be seen as a realisation of Braess' paradox~\cite{Braess1968,witthaut2012} since the outage of an additional link is beneficial in terms of the maximum absolute flow in the network for each individual outage. Hence, cascades of failures may in some situations be prevented by the intentional removal of a second, carefully chosen link after a first transmission link failure threatens stability~\cite{motter_cascade_2004,Witthaut2015}.

\subsubsection{Sign inversion through double outages}
\begin{figure}[t!]
    \begin{center}
       \includegraphics[width=\columnwidth]{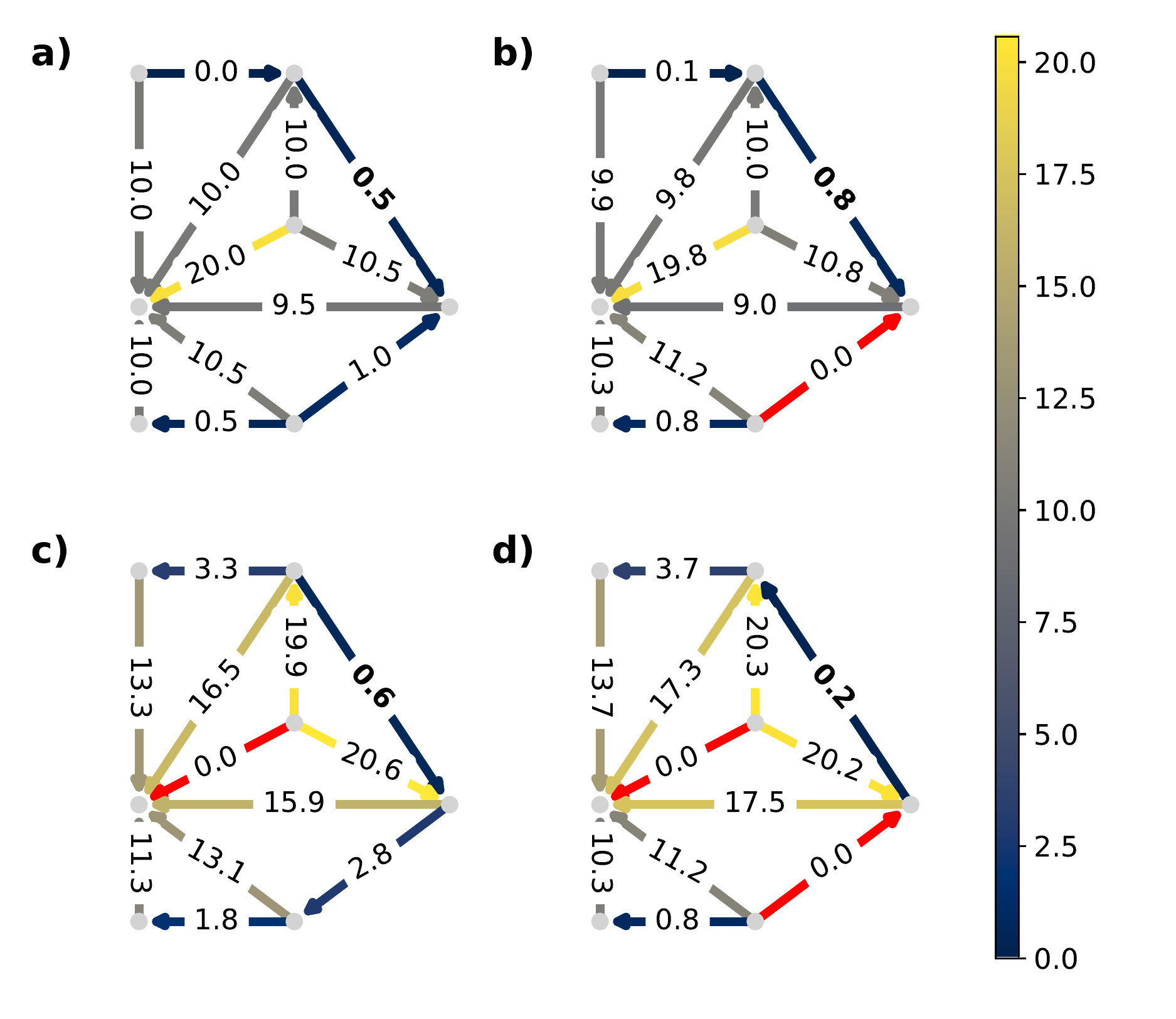}
    \end{center}
    \caption{Collective effects can lead to a complete reversal of the flow changes compared to individual outages. Colour code on lines indicates the magnitude of flow going from blue for no flow to yellow for maximal flow. Red indicates failing links. (a) Initial flow setup with $F_k^{(0)}=20\gg F_o^{(0)}=1$. (b),(c) Flow setup after individual failure of two links ($o$ and $k$ respectively, marked red). In both cases, the flow on the top right link ($l$, bold font) is greater than in the unperturbed grid; $\Delta F_l^{[o]}\approx 0.3$, $\Delta F_l^{[k]}\approx 0.1$. (d) Flow setup after simultaneous failure of both links. The flow on the top right link is smaller than in the unperturbed grid. In fact, not only does the flow change display a sign reversal, the total flow direction is reversed, too; $\Delta F_l^{[o,k]} \approx -0.7 < - F_l^{(0)}\approx -0.5$.}
        \label{fig:collective_cancelling}
\end{figure}
In this section, we will discuss a highly surprising phenomenon that appears in the case of multiple interacting outages; the collective effects may dominate in such a way that purely collective effects can cancel (as shown in the previous section) or even overcompensate the direct effects of individual link failures such that the flow changes resulting from the failure of both links have a different sign compared to the flow changes after each individual failures.

To study this in more detail, we will use the following notation in this section; suppose that links $o$ and $k$ fail and flow changes are monitored on link $l$. Then we denote by $\Delta F_l^{[o]}$ or $\Delta F_l^{[k]}$ the flow changes on link $l$ when link $o$ or $k$ fail, respectively. For the simultaneous outage of both links $o$ and $k$, we denote by $\Delta F_l^{[o,k]}$ the actual flow change on link $l$. With this notation at hand, we will construct examples where flow changes caused by individual link failures $\Delta F_l^{[o]}$ and $\Delta F_l^{[k]}$ have the same sign, but the collective flow change $\Delta F_l^{[o,k]}$ has the opposite sign.

A small example of a network where such a situation occurs is shown in Fig.~\ref{fig:collective_cancelling}. In the initial setup, there is a small flow $F_l=0.5$ on link $l$ (a, bold face number). For the failure of two individual links $o$ and $k$ shown in panels b and c, respectively (red links), the flow on this link is amplified thus showing positive flow changes in both situations $\Delta F_l^{[o]}=0.3$ and $\Delta F_l^{[k]}=0.1$. However, if both links fail simultaneously, the overall flow change has the opposite sign $\Delta F_l^{[o,k]}=-0.7$ thus even inverting the direction of the flow $F_l=-0.2$ with respect to both, the individual setup and the situation after the failure of each individual link.%

We will now explain this surprising phenomenon on a theoretical level in more detail. For simplicity, let us assume the flow changes due to the individual outages to be positive and the flow change in the case of a simultaneous outage to be negative,
\begin{align}
	\Delta F_l^{[o,k]}<0, \qquad \Delta F_l^{[k]}>0,\qquad \Delta F_l^{[o]}>0. \label{eq:weird_case_conditions}
\end{align}
Plugging in Equation~\eqref{eq:flow_changes_davis}, we can cast these three conditions into the following form based on LODFs
\begin{align*}
	1)~ &L_{l,o}L_{o,k}F^{(0)}_k + L_{l,k}L_{k,o}F^{(0)}_o < -(L_{l,k}F^{(0)}_k + L_{l,o}F^{(0)}_o),\\
	2)~ &L_{l,k}F^{(0)}_k>0 \quad \text{ and } \\
	3)~ &L_{l,o}F^{(0)}_o>0.
\end{align*}
To study this condition in detail, we assume that both the initial flows on the failing links and the LODFs between the failing links and the reference link are positive, $F^{(0)}_o, F^{(0)}_k, L_{l,o}, L_{l,k}>0$, without loss of generality \textendash ~this can always be accomplished by redefining the orientation of one or both of the initial flows. We can then immediately deduce that the mutual LODFs both need to be negative $L_{o,k},L_{k,o}<0$ since they always have the same sign (see Appendix~\ref{sec:symmLODF}) and the left-hand-side is positive if these LODFs are positive. Additionally, we do not consider cases where both of the LODFs $L_{k,o}$ and $L_{o,k}$ are equal to (minus) one keeping $\mathcal{L}(o,k)$ finite. For notational convenience, let us now introduce positive constants $\alpha$ and $\beta$ defined by the following quotients;
\begin{align*}
	\alpha(o,k) &= F^{(0)}_k/F^{(0)}_o>0,\\
	\beta(l,o,k) &= L_{l,k}/L_{l,o}>0.
\end{align*}
We thus incorporated the whole dependence on the link monitoring the flow changes $l$ into the purely topological constant $\beta$, whereas any dependence on the flows, i.e. the specific power injections, are incorporated into $\alpha$. Dividing the first condition 1) by the right-hand-side, we arrive at the following inequality%
\begin{align}
	 \frac{\alpha(o,k) |L_{o,k}| + \beta(l,o,k) |L_{k,o}|}{1+\alpha(o,k)\beta(l,o,k)} &> 1. \label{eq:ineq}
\end{align}
In order for the inequality to be fulfilled, we thus need a strong heterogeneity between $\alpha$ and $\beta$, i.e. the ratio of initial flows on links $k$ and $o$ needs to differ strongly from the ratio of their LODFs with respect to link $l$, and we need mutual LODFs $L_{o,k}$ and $L_{k,o}$ that are large in amount in order to reduce the size of the denominator proportional to the product $\alpha\beta$ compared to the numerator. We will see in the next sections that strong mutual LODFs also imply strong collective effects caused by the simultaneous failure of links $k$ and $o$.

Condition~\eqref{eq:ineq} can be simplified drastically if $\alpha\gg\beta$ which may be realised e.g. through a very small initial flow on link $o$ compared to link $k$, such that $F_o^{(0)}\ll F_k^{(0)}$. The above inequality then reduces to
\begin{align*}
	&\frac{\alpha(o,k) |L_{o,k}| + \beta(l,o,k) |L_{k,o}|}{1+\alpha(o,k)\beta(l,o,k)} \approx \frac{|L_{o,k}|}{\beta(l,o,k)},\\
	\Rightarrow &L_{l,o}|L_{o,k}|  > L_{l,k}.
\end{align*}
We will now demonstrate how to design a network where this inequality is satisfied. The construction works as follows; we design a network topology where two links $o$ and $k$ influence each other heavily (measured in terms of LODFs), while a third link $l$ is influenced very differently by each of the links. Formally, $L_{o,k}$ and $L_{k,o}$ both need to be comparatively large, while $L_{l,o}$ and $L_{l,k}$ should be very different in size, thus leading to a small value of $\beta$. In order to further construct an example of flow sign reversal, we choose the power injections $\bm{P}$ such that the flow on link $k$ is much larger than the one on link $o$ such that $\alpha\gg \beta$ and such that $\Delta F_l^{[o]}>0$. The resulting network conforms to the three conditions on the flow changes given in the inequalities~\eqref{eq:weird_case_conditions}.

Indeed, we see that we can find networks where the inequalities are fulfilled as shown in Figure~\ref{fig:collective_cancelling}. The parameters in this case are given by $\alpha(o,k)=20$, $\beta(l,o,k)\approx0.03$, $L_{o,k}\approx-0.19$ and $L_{k,o}\approx-0.23$. Inequality~\eqref{eq:ineq} then holds and reads as
\begin{align*}
    \frac{20\cdot0.19+0.03\cdot0.23}{1+20\cdot 0.03}=2.38 &> 1.
\end{align*}
As discussed previously, the mutual LODFs $L_{o,k}$ and $L_{k,o}$ are both relatively large in this case which is indicative of strong collective interactions as illustrated in the following sections. The purely collective effects not only overshadow the individual outages' effects on link $l$, indeed they reverse the total flow over the link.

\section{Collective effects in complex networks}

\label{sec:coll_effects}
\begin{figure}[t!]
    \begin{center}
       \includegraphics[width=\columnwidth]{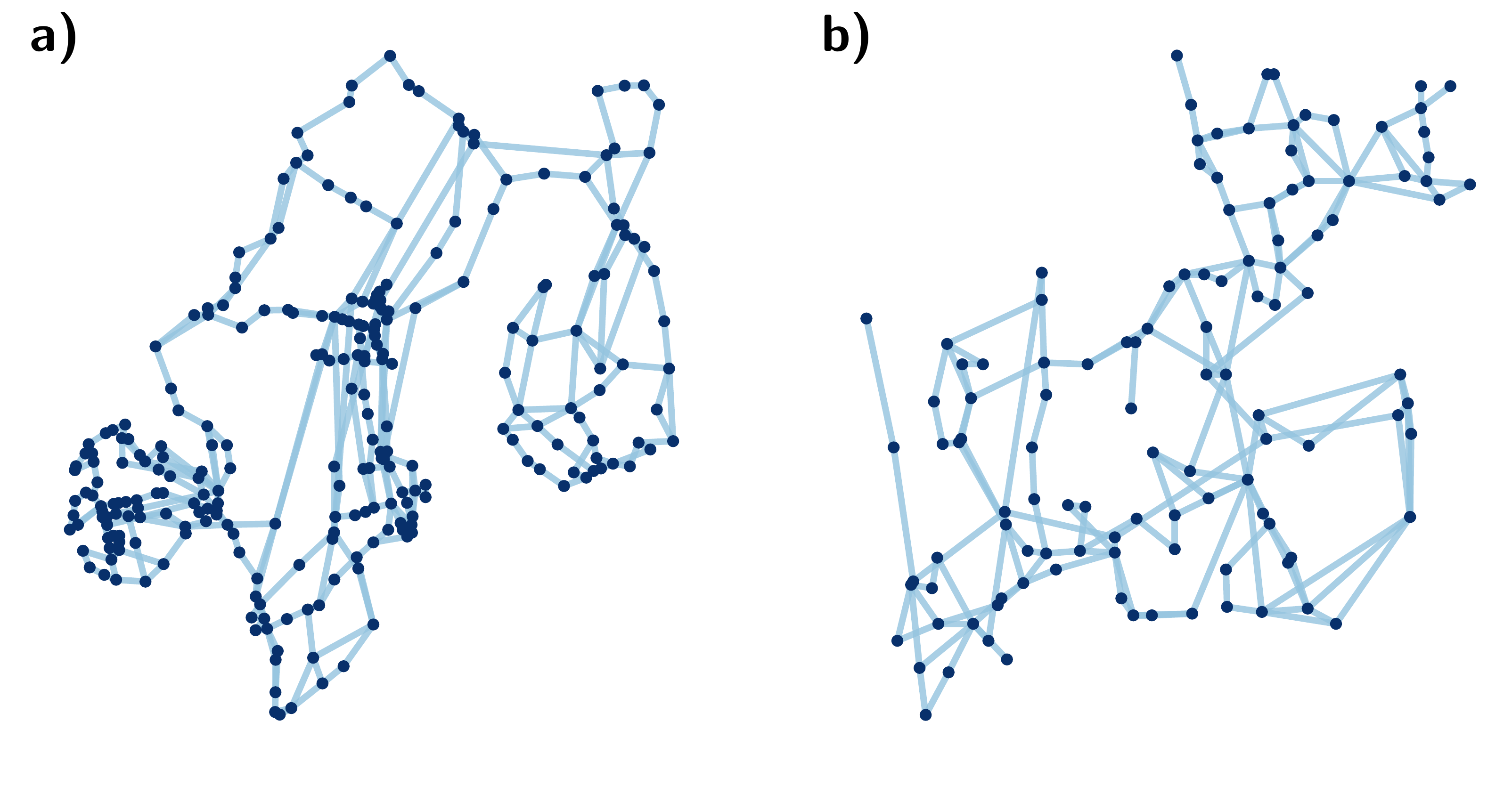}
    \end{center}
    \caption{Two different network topologies are used to demonstrate the performance of the predictor for collective effects. a) Topology of the Scandinavian power grid extracted from an aggregated version of the PyPSA-eur model~\cite{horsch_pypsa-eur_2018} after the removal of dead ends. The resulting topology has 260 nodes and 361 edges. b) Topology of the IEEE test case 118 designed for testing power flow algorithms~\cite{MATPOWER}. The topology has 118 nodes and 179 edges. 
    }
        \label{fig:topologies}
\end{figure}

As shown above, the impact of a double link failure is not given by the simple sum of the individual outages' effects, but strong collective effects may be present. Based on the intuition developed in the last section, we will introduce a quantifier in the following section that may be used to identify in which situations collective effects need to be taken into account and in which situations they may be neglected, thus being able to rely on results obtained for single link failures. We will test this predictor on different test grids, mainly on the ones shown in Figure~\ref{fig:topologies}; the Scandinavian power grid extracted from the software package PyPSA-eur~\cite{horsch_pypsa-eur_2018} (panel a) and the IEEE test case 118~\cite{MATPOWER} (panel b).

\subsection{Quantifying the strength of collective effects}
\begin{figure*}[tb]
    \begin{center}
        \includegraphics[width=\textwidth]{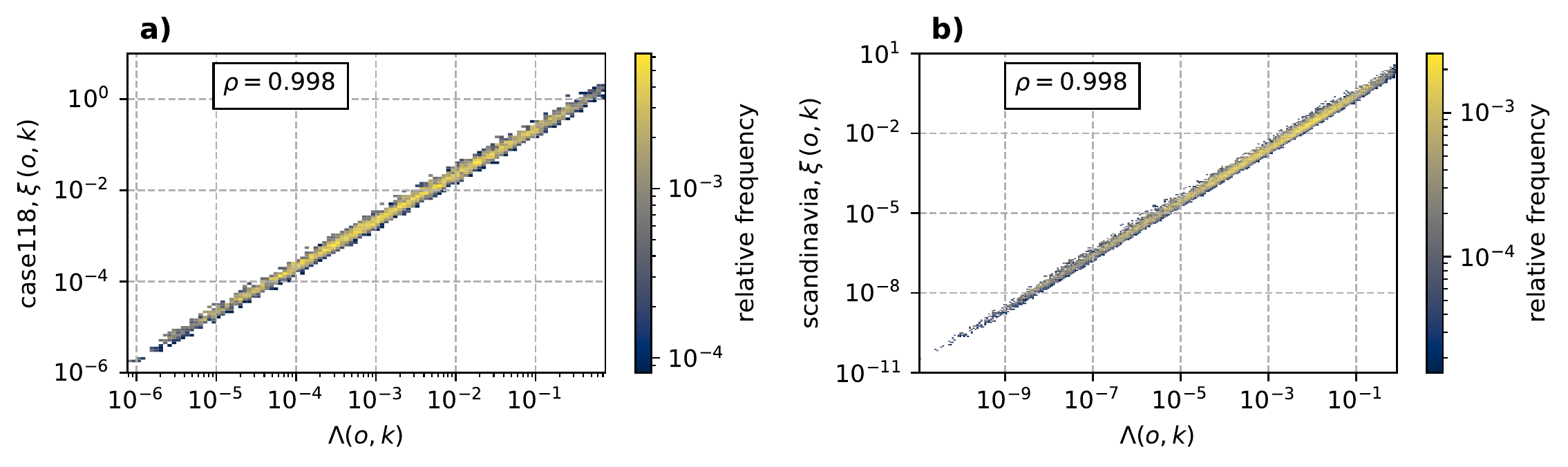}
    \end{center}
    \caption{
    The predictor $\Lambda(o,k)$ performs very well in forecasting the collectivity parameter $\xi(o,k)$ for a link failure of two links $o$ and $k$. In both the IEEE test grid `case118' (a) and the Scandinavian grid (b) the relationship between collectivity parameter $\xi(o,k)$ (ordinate) and predictor $\Lambda(o,k)$ (abscissa) appears to be linear when plotted on a log-log scale. The slope of the curve indicates a linear relationship on the normal scale as well. This implies a strong correlation between the two quantities as implied also by a very large %
    Pearson correlation coefficient of $\rho = 0.998$ in both cases indicating a linear relationship. The histograms' colour code indicates the relative frequency of data points in the given bin. Double logarithmic plots were used to showcase the consistency of the scaling over many orders of magnitude. Note that binning was also done on a double logarithmic basis, leading to much smaller bins for lower values of $\Lambda$ and $\xi$.
    }
        \label{fig:xi_vs_predictor}
\end{figure*}

To understand the purely collective effects of a simultaneous outage of two given links $o$ and $k$, we first calculate the difference between the real flow changes in case of an outage of both links $\Delta F$ and the naive prediction in terms of the sum of individual flow changes $\Delta F^\text{naive}$. The difference calculated according to equation~\eqref{eq:flow_changes_davis} and equation~\eqref{eq:naive_flowchanges}, reads as
\begin{align*}
    \Delta F_l - \Delta F_l^{\text{naive}}&=\mathcal{L}(o,k) \Big((L_{l,o}+L_{l,k}L_{k,o})L_{o,k},\\
     &(L_{l,k}+L_{l,o}L_{o,k})L_{k,o}\Big)^\top
    \begin{pmatrix}
    F^{(0)}_{k}\\F^{(0)}_{o}
    \end{pmatrix}.
\end{align*}
The overall prefactor $\mathcal{L}(o,k)$ is one if the product $L_{o,k}L_{k,o}$ is zero and tends to infinity as the product approaches one. In order to write this expression more compactly, we introduce the matrix $\Xi:\mathbb{R}^2\rightarrow\mathbb{R}^M$ which has the row vectors
\begin{align*}
    \Xi_l &= \mathcal{L}(o,k) [(L_{l,o}+L_{l,k}L_{k,o})L_{o,k},
     (L_{l,k}+L_{l,o}L_{o,k})L_{k,o}]\\
     &=: \mathcal{L}(o,k) [\Xi_l^{(1)},~\Xi_l^{(2)}].
\end{align*}
This matrix includes the topological properties of the rerouting problem and ignores the initial flows $F_o^{(0)}$ and $F_k^{(0)}$, which are determined by the specific power injections which may be time varying. The approach thus allows quantify the impact of collective effects purely based on the network topology. However, this comes at the price of potentially missing situations with very unusual flow patterns in which the approach presented here might not be valid any more to predict collective effects. To get an overall measure of the purely collective part of the failure of two specific lines, we first take the $\ell^2$-norm $\lVert \cdot\rVert_2$ of each row. The resulting vector has the following entry
\begin{align*}
    \lVert\Xi_l\rVert_2&=\mathcal{L}(o,k)\left( \left(\Xi_l^{(1)}\right)^2 + \left(\Xi_l^{(2)}\right)^2 \right)^{1/2} \text{ at position $l$.}
\end{align*}
Taking the $\ell^2$-norm again leads to an overall expression for the purely collective effects of the simultaneous outage of links $o$ and $k$ which we summarise in a single collectivity parameter $\xi$
\begin{align}
    \xi(o,k) &:= \mathcal{L}(o,k) \left( \sum_{l=1}^M \left[ \left(\Xi_l^{(1)}\right)^2 + \left(\Xi_l^{(2)}\right)^2 \right] \right)^{1/2}
    \label{eq:collectivity_parameter}.
\end{align}
Since interpreting the collectivity parameter $\xi$ in this form is rather cumbersome and we are looking for an easily accessible criterion that tells us which pairs of links interact strongly, we further reduce this expression by making a few approximations. Since the LODFs are bounded by one, $-1 \leq L_{a,b} \leq 1$ for all links $a,b$, and are typically much smaller than one, in the order of $\mathcal{O}(10^{-3})$, we expect terms of third order in LODFs to be negligible against terms of second order such that we can on average neglect the former ones. In doing so, we arrive at the following approximation for the collectivity parameter
\begin{align*}
    &\xi(o,k) \approx \mathcal{L}(o,k) \left( \sum_{l=1}^M \left[ \left(L_{l,o}L_{o,k}\right)^2 + \left(L_{l,k}L_{k,o}\right)^2 \right] \right)^{1/2}\\
&=\mathcal{L}(o,k) \left( (L_{o,k})^2 \sum_{l=1}^M \left(L_{l,o}\right)^2 + (L_{k,o})^2\sum_{l=1}^M \left(L_{l,k}\right)^2 \right)^{1/2}.
\end{align*}
Summing over all links in a large network, $L_{l,o}$ and $L_{l,k}$ will vary a lot and may thus essentially be treated as random variables. 
Based on this observation we try to further approximate the above expression. We expect the collectivity parameter $\xi(o,k)$ to be predicted by the two non-varying quantities $L_{o,k}$ and $L_{k,o}$ characterising the interaction between the two failing links. Since LODFs are in general non-symmetric (see Appendix~\ref{sec:symmLODF}), both $L_{o,k}$ and $L_{k,o}$ need to be incorporated to successfully predict the collectivity parameter $\xi(o,k)$. In addition to that, we expect the prefactor $\mathcal{L}(o,k)=(1-L_{o,k}L_{k,o})^{-1}$ to be well approximated by one in general, $\mathcal{L}(o,k)\approx 1$ since (absolute) LODFs are typically small. %

Based on these considerations, we introduce a parameter that predicts the overall strength of collective effects $\xi$ and is defined as follows,
 \begin{align}
     \Lambda(o,k) = \sqrt{L_{o,k}L_{k,o}}.
     \label{eq:predictor}
 \end{align}
 This predictor takes into account the relative effect of the failing links $o$ and $k$ on one another. It is not only a good predictor for the collectivity parameter $\xi(o,k)$, but can also be shown to bound it from below as summarised in the following theorem.
\begin{theorem}
\label{prop:lambdaproof}
Consider a connected network where two links $o$ and $k$ with non-vanishing mutual LODFs $L_{o,k},L_{k,o}\neq 0$ fail. Then the collectivity parameter $\xi(o,k)$ as defined in Equation~\eqref{eq:collectivity_parameter} is bounded from below by the predictor $\Lambda(o,k)=\sqrt{L_{o,k}L_{k,o}}$ 
\begin{align*}
    \xi(o,k)\geq \Lambda(o,k).
\end{align*}
\end{theorem}
A proof is given in Appendix~\ref{sec:proof}. Note that the proof makes use of the fact that $L_{o,o}=-1$ for all links $o$, but we expect the statement to hold even without this assumption. 
Figure~\ref{fig:xi_vs_predictor} illustrates the performance of the predictor in forecasting collective effects for the IEEE test case 118 (panel a) and the Scandinavian power grid (panel b) when averaging over all possible trigger links. The predictor $\Lambda(o,k)$ (abscissa) has a Pearson correlation coefficient with the collectivity parameter $\xi$ (ordinate) of $\rho=0.998$ for both grids, thus indicating a linear relationship between the two quantities. %

The predictor performs equally well if we replace the Euclidean $\ell^2$-norm in the definition of the collectivity parameter $\xi$ in Equation~\eqref{eq:collectivity_parameter} by other $\ell^p$-norms. Norms with $p>2$ tend to emphasise large values much more than smaller norms which is why we also tested the $p=10$-norm and even up to the $p=\infty$-norm, which simply takes the maximum value. The predictor performs very well in predicting collective effects also for other test grids and norms as summarised in table~\ref{tab:predictor_different_norms}. For all norms and all grids tested, we observe a very strong correlation between predictor and collectivity parameter, exceeding $\rho = 0.9$ in most cases. We discuss the predictor and the different norms used to calculate it in more detail in Appendix~\ref{sec:predictor_remainder}. 
 
To summarise, we find that two links show the strongest collective interaction if their mutual LODF values are large, thus implying that a failure of one link has a strong effect on the flow going over the other one and vice versa. 

\begin{table}[tb]
    \centering
    \begin{tabular}{lcccc}
        test grid & \multicolumn{4}{c}{Pearson correlation $\rho$ $\Lambda(o,k)$ vs.\ $\xi(o,k)$} \\
        \hline
         & 1-norm & 2-norm & 10-norm & $\infty$-norm \\
         \hline
         case30 & 0.959&0.98&0.951&0.946\\
         case118 & 0.947&0.984&0.972&0.97\\
         scandinavia &0.909&0.964&0.968&0.967 \\
         pegase1354 &0.933&0.974&0.967&0.966 \\
         Square grid& 0.994 & 0.999 & 0.995 & 0.992\\
         Sparse square grid & 0.946 & 0.968 & 0.981 & 0.976
    \end{tabular}
    \caption{Pearson correlation $\rho$ between predictor $\Lambda(o,k)$ and collectivity parameter $\xi(o,k)$ in the case of a double outage of links $o$ and $k$ for all possible pairs of inks $o$ and $k$ and different test grids. Values are given for a number of test grids, namely IEEE 'case30', 'case118' and 'pegase1354'~\cite{MATPOWER,josz_ac_2016} as well as the Scandinavian grid and a periodic square grid with $20\times 20$ nodes and another one with a share of $s=0.45$ of its links removed, and different norms used to calculate the collectivity parameter $\xi(o,k)$. While $\xi(o,k)$ is predicted very well for all norms, the 2-norm consistently yields the best results, albeit by a small margin.
    \label{tab:predictor_different_norms}
    }
\end{table}

\subsection{Impact of network distance}
\label{sec:impact_distance}
\begin{figure*}[tb]
    \begin{center}
        \includegraphics[width=\textwidth]{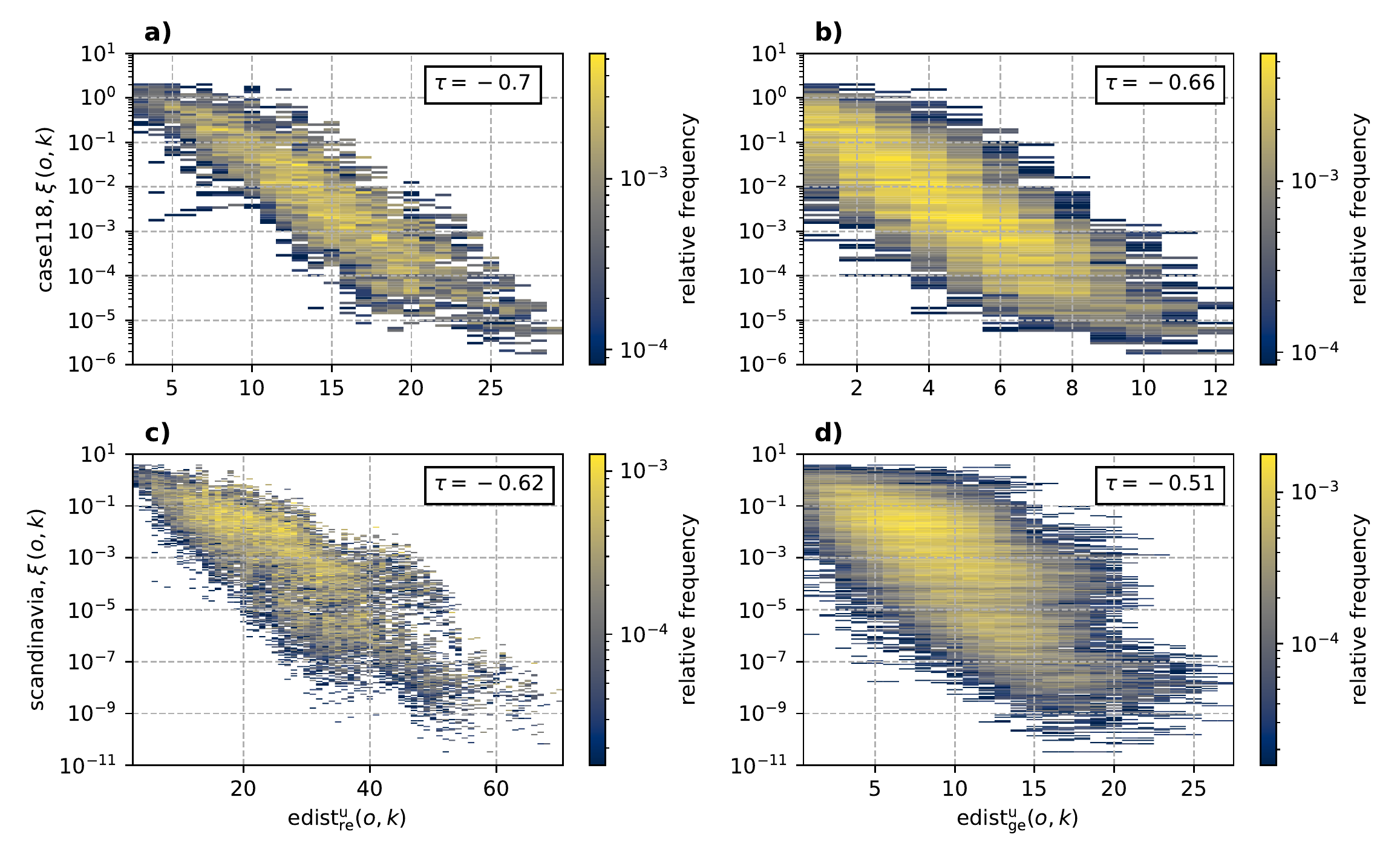}
    \end{center}
    \caption{
    Distance performs moderately in predicting the overall collective effects of a double link failure of two links $o$ and $k$. In both the IEEE test grid `case118' (a,b) and the Scandinavian grid (c,d) the collectivity parameter $\xi(o,k)$ is plotted against the unweighted rerouting distance $\text{edist}_\text{re}^\text{u}(o,k)$ (a,c) and unweighted geodesic distance $\text{edist}_\text{ge}^\text{u}(o,k)$ (b,d), respectively. The Kendall rank correlation $\tau$ is given in all cases. Although there is a clear trend towards smaller collectivity parameters for larger distances, the correlation is much smaller than for the predictor $\Lambda(o,k)$, thus indicating that effects other than distance play an important role for collective effects as well.%
    The histograms' colour code indicates the relative frequency of data points in the given bin. Logarithmic plots were used to resolve more details for very small values of the collectivity parameter.
    Note that binning was also done on a logarithmic basis, leading to much smaller bins for lower values of $\xi$.}
    \label{fig:xi_vs_distances}
\end{figure*}
Distance is known to play an important role for failure spreading in power grids and other types of linear flow networks~\cite{strake2018non,Kett15,Labavic2014,Jung2015}. In this section, we will examine if they may also be used to successfully predict collective effects in multiple link failures. Typically, distances in networks are measured between two nodes with the most prominent distance measure being the geodesic distance. It is given by the sum of the lengths or weights of all edges along a shortest path between the respective nodes,
\begin{align*}
    \text{dist}^\text{u/w}_0(v_1,v_2) &= \min_{\text{paths } p(v_1,v_2)} \sum_{e\in p} \ell_e,
\end{align*}
where the superscript `u' or `w' denotes the unweighted or weighted distance, the subscript `0' describes the distance in the initial graph before any kind of outage, $v_1$ and $v_2$ are the nodes whose distance is calculated, $p(v_1,v_2)$ is a path from $v_1$ to $v_2$ and $\ell_e$ is the length or weight of edge $e$, which is set to unity when calculating unweighted distances. Additionally, one can define the geodesic distance between edges as the smallest possible distance between the nodes incident to the corresponding edges plus half of each edge's length,
\begin{align*}
    &\text{edist}^\text{u/w}_\text{ge}[(r,s),(m,n)] \\
    &= \min_{v_1\in\{r,s\} , v_2 \in \{m,n\} } {\rm dist}_0^\text{u/w}(v_1,v_2)+\frac{\ell_{(r,s)}+\ell_{(m,n)}}{2}. 
\end{align*}
Here the subscript `ge' denotes the geodesic distance while $(r,s)$ and $(m,n)$ are the respective edges given by the nodes they are incident on. As we demonstrated in a recent publication~\cite{strake2018non}, this distance measure does not capture essential aspects of the flow rerouting after a link failure. Instead, we proposed the rerouting distance 
\begin{align*}
    \text{edist}_\text{re}^{u/w}[(r,s),(m,n)],
\end{align*} 
given by the length of the shortest cycle crossing both edges $(r,s)$ and $(m,n)$. If no such path exists, the rerouting distance is defined to be $\infty$. This distance measure is strongly correlated with the magnitude of the LODFs as shown in Ref.~\cite{strake2018non}.

Figure~\ref{fig:xi_vs_distances} shows the scaling of $\xi$ with distance between the failing links for a failure of two links. Here, we make use of the Kendall rank correlation $\tau$ to quantify the degree of correlation between the two quantities which quantifies the rank correlation. This is due to the fact that in contrast to the scaling observed with the predictor $\Lambda(o,k)$ observed in the last section, we do no observe a linear scaling of the collectivity parameter $\xi$ with the different distance measures. The rerouting distance performs slightly better in predicting the collectivity parameter $\xi$ than the geodesic distance, where the former one has a rank correlation of $\tau=-0.7$ and $\tau=-0.62$ with $\xi$ and the latter one a correlation of $\tau=-0.66$ and $\tau=-0.51$ in the test grid 'case118' and the Scandinavian grid, respectively. Thus both distance measures perform moderately in predicting the collectivity parameter although not nearly as well as the predictor introduced in the last section. Still, the distance seems to be an important factor in determining the simultaneous outages' effects \textendash ~but contrary to the case of a single outage~\cite{strake2018non}, other factors play an important role, too. We may thus deduce that links that are closer to each other in both, the rerouting distance and the simple edge distance tend to have a stronger collective response. This behaviour is expected given that the predictor performing best is given by the product of the mutual LODFs between the two failing links and the rerouting distance is known to be a good predictor for the LODF~\cite{strake2018non}.

\section{Extension to arbitrary link failures}
\label{sec:arbitrary_failures}
Now that we analysed the simultaneous failure of two links in detail, we will extend the theoretical framework to more than two links failing. To this end, we will derive a formula that describes this type of contingencies on a nodal level and perform a continuum limit for valid for infinitely large regular grids.

\subsection{Derivation of generalized LODFs}
Now consider the simultaneous outage of $K$ links $\{l_1,...,l_K\}$ with $K<M$. Then we define the projection matrix from the space of all links onto the subset of failing links $\boldsymbol{\mathcal{P}}:\mathbb{R}^M\rightarrow\mathbb{R}^K$ via
\begin{align*}
    \mathcal{P}_{kl} &= \delta_{l,l_k},
\end{align*}
where $\delta_{l,l_k}$ denotes the Kronecker delta. Consequently, let $\vec D: \mathbb{R}^K\rightarrow\mathbb{R}^N$ be the projection of the node-edge-incidence matrix $\bm I:\mathbb{R}^M\rightarrow\mathbb{R}^N$ onto the subset of failing links which reads as
\begin{align*}
    D_{nk} &= (\bm I\boldsymbol{\mathcal{P}}^\top)_{nk} = I_{n,l_k},\text{ for } k \in \{1,2,\ldots,K\}.
\end{align*}
With this definition the columns of this matrix are the vectors $\bm d_{l_k}$ introduced in the definition of the LODF, as per Eq.~\eqref{eq:LODF}. As a reminder, they are defined by their entries being $+1$ at the node corresponding to the start of the respective failing link, $-1$ at the node corresponding to the end of the failing link and $0$ otherwise. 
Furthermore, we define the projected branch reactance matrix $\vec B_\mathrm{out}:\mathbb{R}^K\rightarrow\mathbb{R}^K$ by
\begin{align*}
    \vec B_\mathrm{out} &= \boldsymbol{\mathcal{P}}\vec B_\mathrm{d}\boldsymbol{\mathcal{P}}^\top = \mathrm{diag}(b_{l_1}, b_{l_2}, \ldots, b_{l_K}).
\end{align*}
Using this matrix, we can also project the vector of all initial flows onto the failing links defined by
\begin{align*}
    \vec F^{(0)}_\mathrm{out} &= \boldsymbol{\mathcal{P}}\vec F^{(0)} =\vec B_\mathrm{out} \vec D^\top \vec\theta
        = (F^{(0)}_{l_1}\ F^{(0)}_{l_2}\ \cdots\ F^{(0)}_{l_K})^\top.
\end{align*}
The failure of multiple links may be regarded as a perturbation to the graph Laplacian $\vec B$ in the same way as for a single link, see Ref.~\cite{strake2018non}
\begin{align*}
    \vec{\hat{B}}=\bm{B}+\Delta\bm{B},
\end{align*} 
where $\vec{\hat{B}}$ is the graph Laplacian after the failure of the $K$ links. The corresponding perturbation matrix $\Delta \vec B$ may then also be expressed using the projected node-edge-incidence matrix as
\begin{align}
    \Delta \vec B &= -\vec D \vec B_\mathrm{out} \vec D^\top. \label{eq:perturbation_matrix}
\end{align}
In addition to that, the failure causes the nodal potentials to change
\begin{align*}
    \hat{\bm{\theta}}=\bm{\theta}+\bm{\psi},
\end{align*}
where $\bm{\psi}$ is a vector of the changes in angles. Using the continuity equation~\eqref{eq:DCapprox} in the new grid
\begin{align*}
    \bm{P}=(\bm{\theta}+\bm{\psi})(\bm{B}+\Delta\bm{B}),
\end{align*}
subtracting from it the current balance for the old grid, and applying the Moore-Penrose-pseudoinverse to the resulting equation, the change in potential is calculated as 
\begin{align}
    \bm{\psi}=-(\bm{B}+\Delta\bm{B})^\dagger\Delta\bm{B}\bm{\theta}.
    \label{eq:angle_change}
\end{align}
We can simplify this expression by making use of the Woodbury matrix identity~\cite{Wood50} and arrive at the final result
\begin{align}
    \vec\psi
    &= \vec B^\dagger\vec D (\bm{1}_K - \mathfrak{P})^{-1}\vec F_\mathrm{out}^{(0)}. \label{eq:phase_angle_changes}
\end{align}
Here, we defined a projection of the PTDF matrix onto the subset of failing links $\mathfrak{P}:\mathbb{R}^K\rightarrow\mathbb{R}^K$ given by
\begin{align*}
    \mathfrak{P} &:= \vec B_\mathrm{out} \vec D^\top \vec B^\dagger \vec D = \boldsymbol{\mathcal{P}}\ \mathbf{PTDF}\ \boldsymbol{\mathcal{P}}^\top.
\end{align*}
The change in phase angles may then be used to calculate the flow changes by making use of Equation~\eqref{eq:flow-line}. The vector of flow changes reads as 
\begin{align}
    \Delta \vec F  %
        &= \vec B_\mathrm{d}\vec I^\top\vec B^\dagger\vec D (\bm{1}_K - \mathfrak{P})^{-1}\vec F_\mathrm{out}^{(0)}. \label{eq:flow_changes}
\end{align}
 In principle, we may now make use of Eq.~\eqref{eq:flow_changes} to calculate the flow changes after an arbitrary number of simultaneous contingencies. The immediate insight into the structure of the contingency problem from this equation is, however, limited. We will thus try to gain more insight into the interplay of multiple outages by rearranging the equation.

Starting with Equation~\eqref{eq:phase_angle_changes} expressing the change of voltage phase angles after the failure $\vec{\psi}$, we can derive the following Poisson-like equation similar to the case of a single link failure as presented in Equation~\eqref{eq:Poisson}
\begin{align}
    \bm B \vec\psi &= \vec D \vec F^\mathrm{(K)},
    \label{eq:field_unfinished}
\end{align}
where we defined the vector of flows weighted by the dipole source terms
\begin{align*}
    \vec F^\mathrm{(K)}:=(\bm{1}_K - \mathfrak{P})^{-1} \vec F^{(0)}_\mathrm{out}.
\end{align*}
We may thus rewrite Equation~\eqref{eq:field_unfinished} for the change in nodal potentials as follows, making the correspondence to the Poisson equation more apparent;
\begin{align}
    \vec B \vec\psi &= \sum_{k=1}^K \vec q_k, \label{eq:poisson}
\end{align}
with the dipole sources \begin{align*}
    \vec q_k = \vec d_{k} F_k^\mathrm{(K)}
\end{align*} 
and the $\bm{d}_k$ being the rows of $\bm{D}$, see also section~\ref{sec:single_and_double}. In addition to this expression, we can derive an analogous equation for the graph $\hat{G}$ from which all the failing lines have been removed. Simply plugging Equation~\eqref{eq:perturbation_matrix} into Equation~\eqref{eq:angle_change}, we arrive at the following equation
\begin{align*}
        \vec{\hat{B}} \vec\psi &= \vec{D}\vec{F}^{(0)}_{\text{out}}.
\end{align*}

We are thus left with a \emph{discrete Poisson equation} with potential $\vec\psi$, which is analogous to the result obtained in our previous work~\cite[sections~III-IV]{strake2018non} for a single failing link. Instead of a single dipole source this equation is governed by $K$ dipole sources. However, this equation differs from the naive approach obtained by simply superposing single dipole sources. To see this, consider the case of $K=2$ link failures. As we have seen in the last section~\ref{sec:coll_effects}, collective effects play an important role in the interaction of the two links. In this case, a simple superposition of two dipoles results in the equation 
\begin{align*}
     \vec B \vec\psi^\mathrm{naive} &= \bm{q}_1^{\text{naive}} +  \bm{q}_2^{\text{naive}},
\end{align*}
where we defined the dipole sources resulting from the naive approach
\begin{align*}
    \bm{q}_1^{\text{naive}}&=\vec d_1 (1 - \mathfrak{P}_{11})^{-1} F^{(0)}_{\mathrm{out},1},\\
    \bm{q}_2^{\text{naive}}&=\vec d_2 (1 - \mathfrak{P}_{22})^{-1} F^{(0)}_{\mathrm{out},2}.
\end{align*}
On the other hand, using the exact approach in Equation~\eqref{eq:poisson}, the actual dipole sources read as
\begin{align*}
     \bm{q}_1 &= \vec d_1 \left([(\bm{1}_2 - \mathfrak{P})^{-1}]_{11} F^{(0)}_{\mathrm{out},1} + [(\bm{1}_2 - \mathfrak{P})^{-1}]_{12} F^{(0)}_{\mathrm{out},2}\right),\\
     \bm{q}_2 &=\vec d_2 \left([(\bm{1}_2 - \mathfrak{P})^{-1}]_{21} F^{(0)}_{\mathrm{out},1}+(\bm{1}_2 - \mathfrak{P})^{-1}]_{22} F^{(0)}_{\mathrm{out},2}\right).
\end{align*}
The naive approach thus underestimates the interaction between the two dipole sources encoded in the matrix inverse $(\bm{1}_2 - \mathfrak{P})^{-1}$. We discussed this interaction in detail in Section~\ref{sec:coll_effects}. In the following paragraph, we will demonstrate, however, that this collective effect can be neglected in the continuum limit, thus making the naive superposition approach exact in that case.

\subsection{Continuum limit for regular square lattice}
We will now demonstrate how one may derive an exact formula for the potential changes after an arbitrary number of link failures for the setup of an infinite square lattice extending on our previous work~\cite{strake2018non}. Consider the elementary example of a regular square lattice embedded in the plane $\mathbb{R}^2$. Label all nodes by their positions $\vec r = (x,y)$ and let the lattice spacing be denoted by $h$. Now introduce continuous functions $\psi$ and $b$ such that $\psi(x,y)$ is the potential of the node at $(x,y)$ and $b(x+h/2,y)$ is the weight of the link connecting the two nodes at $(x,y)$ and $(x+h,y)$ and analogously for two nodes connected in $y$-direction. 

For a small lattice spacing $h\rightarrow 0$ and an infinitely large grid, the left-hand side of the Poisson equation~\eqref{eq:poisson} evaluated at position $(x,y)$ can be written in a continuum version as~\cite{strake2018non}
\begin{align}
    (\vec B\psi)(x,y) &= -h^2\bm\nabla (b(x,y)\bm\nabla\psi) + \mathcal{O}(h^3). \label{eq:LHS}
\end{align}
Then, the flow changes according to Equation~\eqref{eq:flow_changes} are given by
\begin{align*}
    \Delta\vec F(x,y) &= b(x,y)\bm\nabla\psi(x,y),
\end{align*}
where $\Delta\vec F$ refers to the change in flow due to the link failures here and should not be confused with the continuous Laplace operator.

The right-hand side of the Poisson equation~\eqref{eq:poisson} may be calculated similarly noting that at most $2K$ nodes contribute when $K$ links fail. Let any failing link $l_k \in \{l_1, l_2, \ldots, l_K\}$ connect the nodes $s_k$ and $t_k$ with positions $(x_{s_k}, y_{s_k})$ and $(x_{t_k}, y_{t_k})$ respectively.
The discrete version of a single addend on the right-hand side reads:
\begin{align*}
    \vec q_k &= \left[(\bm{1}_K - \mathfrak{P})^{-1} \vec F^{(0)}_\mathrm{out}\right]_k \vec d_{k}\\
    &= \vec d_{k}
        \sum_{i=1}^K [(\bm{1}_K - \mathfrak{P})^{-1}]_{ki} F^{(0)}_{\mathrm{out},i}.
\end{align*}

We will now show how this equation may be interpreted in the continuum version. First, the flow on a failing link before the outage $F^{(0)}_{l_i}$ may be calculated as
\begin{align*}
    F^{(0)}_{l_i}\ \hat{=}\ h\vec F^{(0)}(x_{s_i}, y_{s_i}) + \mathcal{O}(h^2),
\end{align*}
where $\vec F^{(0)}(x_{s_i},y_{s_i}) = b(x_{s_i}, y_{s_i})\bm\nabla\theta(x_{s_i}, y_{s_i})$ is the continuum version of the flow before the outage. Second, the vector $\vec d_{k}$ can be formally interpreted in terms of the two-dimensional delta function $\delta(x,y)$ and reads for a link $l_k$ oriented in $x$-direction
\begin{align*}
    \vec d_{k}\ &\hat{=}\ h\partial_x\delta(x-x_{s_k}, y-y_{s_k}) + \mathcal{O}(h^2).
\end{align*}
For links oriented in $y$-direction, we simply replace $\partial_x$ by $\partial_y$. Finally, in order to calculate the continuum version of the inverse matrix elements $[(\bm{1}_K - \mathfrak{P})^{-1}]_{ki}$ for two arbitrarily chosen links $l_k$ and $l_i$, assume without loss of generality that both links are oriented in $x$-direction and that a continuum version $b^\dagger$ of the Green's function $\bm B^\dagger$ exists. Then, the elements of the projected PTDF matrix may be calculated as
\begin{align*}
    \mathfrak{P}_{ki} =\ & b_k \vec d_k^\top \vec B^\dagger \vec d_i \\
    \ \hat{=}\ & h^2 b(x_{s_k}, y_{s_k}) \int\partial_y\delta(x-x_{s_k},y-y_{s_k}) b^\dagger(x,y)\\ &\partial_x\delta(x-x_{s_i},y-y_{s_i})\,\mathrm{d}x\mathrm{d}y \\
    =\ & \delta_{ki} \left[ h^2 b(x_{s_k}, y_{s_k}) \frac{\partial^2 b^\dagger(x_{s_k}, y_{s_k})}{\partial x\partial y} + \mathcal{O}(h^3) \right].
\end{align*}
 All off-diagonal entries are zero due to the delta functions' different arguments. Importantly, this observation is independent of the orientation of the two links under consideration. The inverted matrix is thus diagonal and can be calculated as
\begin{align*}
    [(\bm{1}_K - \mathfrak{P})^{-1}]_{ki}\ &\hat{=}\ (1- \mathfrak{P}_{ki})^{-1} = (1-\mathcal{O}(h^2))^{-1}
\end{align*}

In total, we obtain after expanding the entire expression to lowest order in the continuum limit
\begin{align}
    q_k(x,y) &= h^2 \vec F^{(0)}(x_{s_i}, y_{s_i})^\top \bm\nabla\delta(x-x_{s_k}, y-y_{s_k}) + \mathcal{O}(h^3). \label{eq:RHS}
\end{align}
We can now formally divide the left-hand side~\eqref{eq:LHS} and the right-hand side~\eqref{eq:RHS} by $h^2$ and take the limit $h\rightarrow 0$ to obtain the final continuum limit of the Poisson equation,
\begin{align}
    \bm\nabla (b(x,y)\bm\nabla\psi) &= -\sum_{k=1}^M \vec q_k^\top\bm\nabla\delta(x-x_{s_k}, y-y_{s_k}),
\end{align}
where the source terms are $\vec q_k(x_{s_k},y_{s_k}) = \vec F^{(0)}(x_{s_k},y_{s_k})$, the unperturbed current field.

If the link weights are homogeneous, $b(x,y)=b$, the solution is given by the superposition of $K$ two-dimensional dipole fields
\begin{align}
    \psi(\vec r) &= \sum_{k=1}^K \frac{\vec q_k (\vec r - \vec r_k)}{|| \vec r - \vec r_k ||^2}, \label{eq:dipoles_potential} \\
    \Delta\vec F(\vec r) &= b\cdot\sum_{k=1}^K\left( 
        \frac{\vec q_k}{|| \vec r - \vec r_k ||^2} - 2(\vec r - \vec r_k) \frac{\vec q_k\cdot(\vec r - \vec r_k)}{|| \vec r - \vec r_k ||^4} 
        \right). \label{eq:dipoles_field}
\end{align}
We thus obtain a fully analytic solution in the continuum limit. This solution reveals that in homogeneous lattices the effects of multiple outages are given by the superposition of single outages.

\section{Conclusion and outlook}
\label{sec:conclusion}
In this article, we showed that multiple link failures can lead to fundamentally different impacts than expected from a naive superposition of single link failures. We also established a parameter quantifying in which cases these effects have to be taken into account. We extended on previous work demonstrating that multiple link failures correspond to the overlay of correspondingly many single dipoles in infinitely large regular grids, thus allowing for a description similar to single link failures in this case. However, the strength of the effective dipoles is strongly determined by the collective effects, i.e. the interplay of the failing links. Our results demonstrate that further understanding of multiple link failures is an important task for the development and security of future power systems, thus helping to understand in which cases additional link shutdowns can help or counteract overall system security.

In this work, we introduced several elementary examples which demonstrate the counterintuitive behaviour of collective effects in some particular cases. We showed that additional outages can be beneficial for the overall grid loading, thus presenting another occurence of Braess' paradox in power grids. On the other hand, we showed that collective effects might lead to a sign inversion of flow direction compared to the individual failure of each single link. Both phenomena are potentially of high relevance when operating power grids as they might help to resolve situations where a single link fails or a redispatch occurs. However, further work should be dedicated to understanding and predicting these particular collective effects on a more fundamental theoretical basis.

The predictor for collective flow changes introduced in this manuscript allows for an easier understanding of when collective effects on the flow changes become considerably important. Mostly, collective effects seem to be small, they only become relevant in cases where both failing links have a strong effect on one another. This is for example the case if the links are in close proximity or if they are both bottlenecks. Conversely, this implies that the intuition developed for single link failures may in many cases also be applied to study multiple link failures if the possibility of collective effects is kept in mind.

Distance between the two failing links seems to play an important role for the overall collective effect. Previous work has addressed the role of distance in flow changes for single link failures where in particular the rerouting distance was shown to be a decisive measure in predicting the flow changes~\cite{strake2018non}. In predicting collective effects, rerouting distance between the failing links still seems to be an important quantity but not to the same degree as it is important for single link failures. In the future, it would be interesting to extend the rerouting distance to more than two links which could potentially also predict collective effects better.

Further work should address the role particular flow patterns play in more detail. In the approach used here, we abstract from individual flow patterns and focused on topological aspects. This should be a good approximation for many cases, in particular when dealing with large power grids. However, there might be cases in which specific lines are nearly always more heavily loaded than other ones which should imply a more important contribution of such lines to collective effects.

\acknowledgments

We gratefully acknowledge support from the German Federal Ministry of Education and Research (BMBF grant no. 03SF0472) and the Helmholtz Association (via the joint initiative ``Energy System 2050 -- a contribution of the research field energy'' and the grant no. VH-NG-1025 to D.W.).

\newpage
\newpage
\appendix

\begin{figure*}[tb!]
\begin{center}
\includegraphics[width=1.\textwidth]{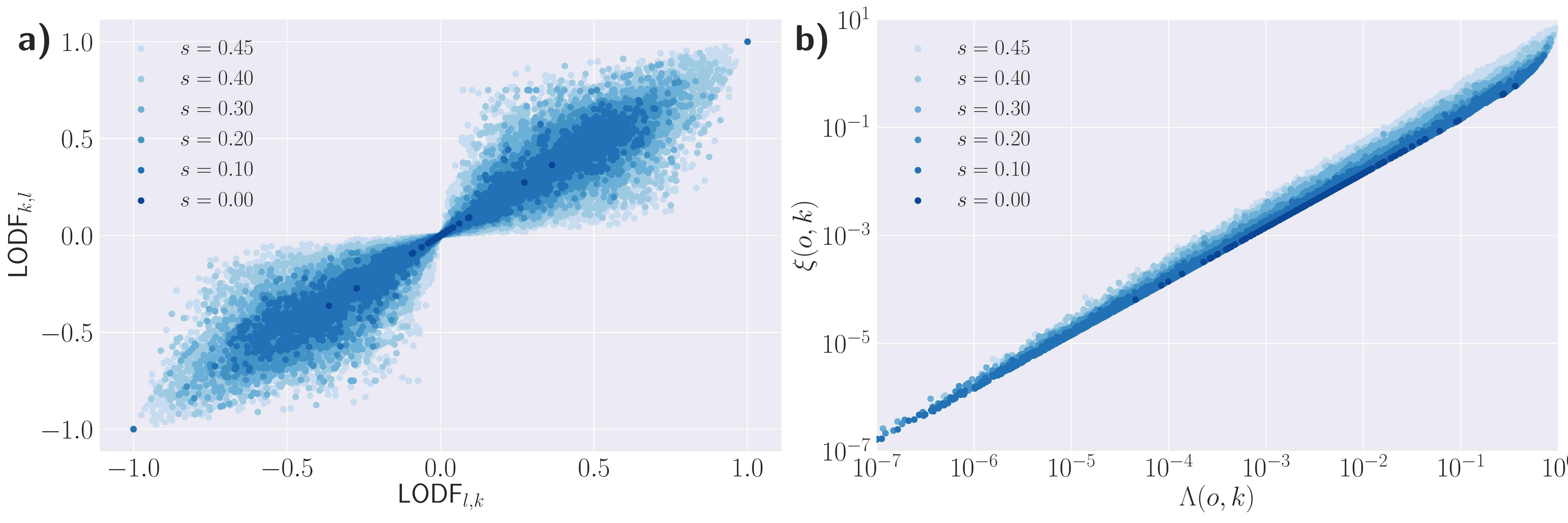}
\end{center}
\caption{
\label{fig:sparse_sg}
With increasing degree of sparsity $s$ in square grids, the LODFs become less symmetric. a) Whereas we observe fully symmetric LODFs in a periodic square grid without any edges removed $s=0$, removing edges increases the degreee of asymmetry continuously up to $s=0.45$. This is due to the fact that the entries of the inverse Laplacian $\bm{B}^\dagger$ become increasingly heterogeneous with more edges removed. b) The degree of asymmetry in LODFs induced by increased sparsity in periodic square grids also influences the performance of the predictor for collective effects $\Lambda(o,k)$; it performs almost perfectly for a periodic square grid with no links removed and homogeneous edge weights, where also LODFs are perfectly symmetric. With increasing degree of sparsity, the performance reduces slightly, see also table~\ref{tab:predictor_different_norms}. For very high values of the predictor, the prefactor $\mathcal{L}(o,k)$ dominates leading to the change from a linear scaling to a non-linear scaling for these values.
}
\end{figure*}
\section{Symmetry of LODFs}
\label{sec:symmLODF}

The LODFs according to Eq.~\ref{eq:LODF} are given by
\bee
    \text{LODF}_{l,k}=b_{l}\frac{\bm d_{l}^\top\bm B^\dagger \bm d_{k}}{1-b_{k}\bm{d}_{k}^\top\bm B^\dagger \bm d_{k}}.
\eee

In this section, we will study the symmetry of this matrix in terms of interchanging the role of failing link $k$ and link where flow changes are monitored $l$. This symmetry describes the extend to which the flow change on one link $l$ due to another, failing link $k$ corresponds to the opposite flow change on link $k$ if link $l$ fails instead and thus provides a measure of symmetry for the whole network. In particular, we analyse how the matrix becomes asymmetric with an increasing degree of asymmetry in the links surrounding the monitored, and failing link. This explains why both LODFs are important for predicting the strength of collective effects in the predictor~\eqref{eq:predictor}.

If we assume homogeneous edge weights for the time being such that $\bm{B} = b\cdot\bm{I}$, we notice that the numerator in Eq.~\eqref{eq:LODF} is symmetric with respect to interchanging $l$ and $k$. This numerator is also referred to as Power Transfer Distribution Factor (PTDF) in power engineering~\cite{Wood14}. The symmetry can be seen by taking the matrix transpose of the expression 
$$(\bm d_{l}^\top\bm B^\dagger \bm d_{k})^\top = \bm d_{k}^\top(\bm B^\dagger)^\top \bm d_{l}=\bm d_{k}^\top\bm B^\dagger \bm d_{l}.$$

On the other hand, the denominator is non-symmetric even in the case that line susceptances are uniform, as it reads as $1-b\cdot\bm{d}_{k}^t\bm B^\dagger \bm d_{k}$ for $\text{LODF}_{l,k}$ and $1-b\cdot\bm{d}_{l}^t\bm B^\dagger \bm d_{l}$ for $\text{LODF}_{k,l}$ thus encoding the importance of the link that fails. The LODFs are only completely symmetric if both links have not only the same weights, but also the same topological structure around them. This is for example the case for the periodic square grid, see Fig.~\ref{fig:sparse_sg}, dark blue dots corresponding to $s=0$ in the legend. We analysed this expression in detail in our previous publication~\cite{strake2018non} and showed that it can be predicted using the minimum cut that disconnects the two vertices $k_1$ and $k_2$. In the case of uniformly distributed line susceptances, this expression simply reduces to the well-studied resistance distance~\cite{Klei93,Xiao2003}. Thus, the asymmetry in LODFs is encoded in both, the asymmetry in connectivity in the network, i.e. the variance in the node degree, and the asymmetry in the line susceptances.

In addition to the observation made before, we can notice that the mutual LODFs $L_{l,k}$ and $L_{k,l}$ will always have the same sign. This is due to the fact that as discussed above, the numerator is the same for both expression. On the other hand, the denominator is always positive or equal to zero 
\bee
1-b_{k}\bm{d}_{k}^\top\bm B^\dagger \bm d_{k}\geq 0.
\eee
Whereas this is not obvious from the above expression, it follows from the definition of the PTDFs given by $b_{k}\bm{d}_{k}^\top\bm B^\dagger \bm d_{k}=\text{PTDF}_{k,k}\in [-1,1]$~\cite{Wood14}. Therefore we may conclude that mutual LODFs always have the same sign.

In Figure~\ref{fig:sparse_sg}, we demonstrate how asymmetry in LODFs arises with an increasing degree in inhomogeneity in the nodal degrees for a periodic square lattice from which we randomly remove an fraction $s$ of its total number of links according to the procedure described in Ref.~\cite{strake2018non}. In this figure, we plot the $\text{LODF}_{l,k}$ against the $\text{LODF}_{k,l}$ for all possible combinations of links $l$ and $k$. Starting at $s=0$ for a perfect periodic square lattice with $50\times 50$ nodes, the LODF is perfectly symmetric (dark blue dots). With increasing degree of sparsity $s\in \{0.1,0.2,0.3,0.4,0.45\}$ (dots from dark blue to light blue), we observe an increasing spread of the LODFs indicating an increasing degree of asymmetry in the LODFs.

\section{A predictor for collective effects}
\label{sec:predictor_remainder}

\begin{figure*}[tb!]
\begin{center}
\includegraphics[width=1.\textwidth]{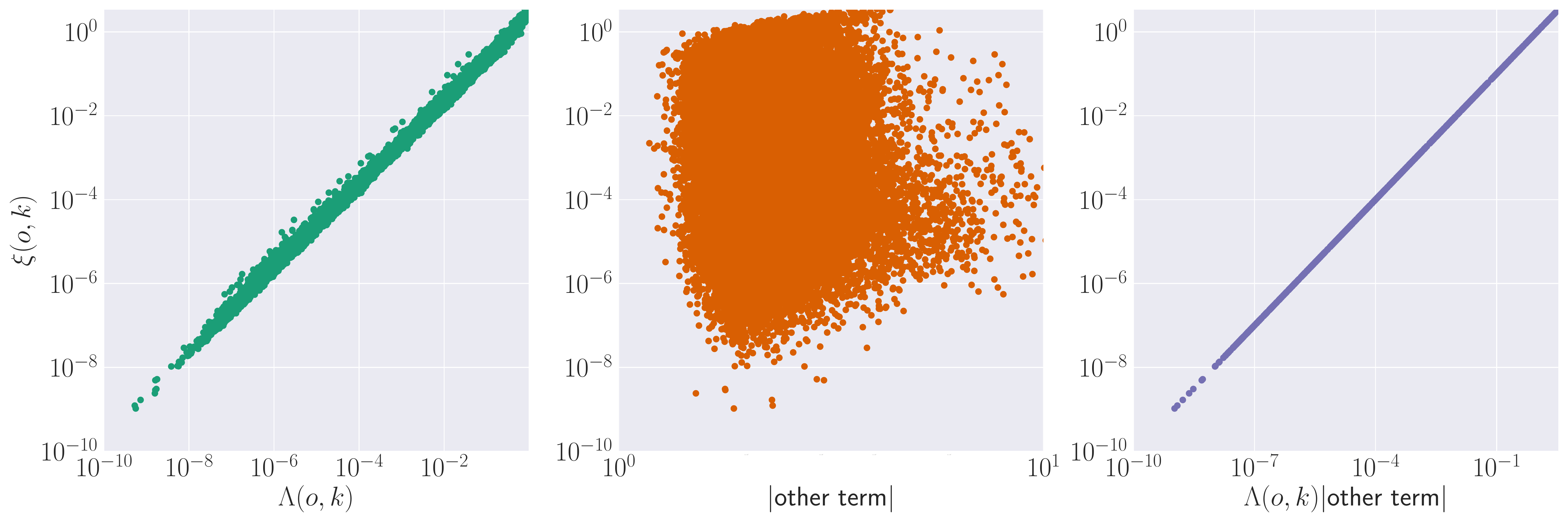}
\end{center}
\caption{
\label{fig:predictor_and_remainder}
(a) The predictor $\Lambda(o,k)$ performs equally well predicting the collectivity parameter $\xi$ for the test case 'pegase1354'~\cite{josz_ac_2016}. Predictor and collectivity parameter are plotted on a log-log-scale. (b) The remaining term does not show any visible correlation with the collectivity parameter $\xi$. (c) The product of both terms exactly reproduces the collectivity parameter $\xi$ as expected.
}
\end{figure*}

To support the choice of the predictor for collective effects, consider Figure~\ref{fig:predictor_and_remainder}(a,b). We show the predictor $\Lambda(o,k)$ and the remaining term, referred to as 'other term' in the figure, for all possible combinations of trigger links for the test case 'pegase1354'~\cite{josz_ac_2016}. The remaining term is constructed by factoring out the predictor $\Lambda(o,k)$ defined in Eq.~\ref{eq:predictor} of the collectivity parameter $\xi(o,k)$ in Eq.~\ref{eq:collectivity_parameter} and assuming both LODFs $L_{k,o}$ and $L_{o,k}$ to be non-zero
\begin{align*} 
    &\xi(o,k) = \Lambda(o,k)\cdot\\
    &\mathcal{L}(o,k)\Bigg(\sum_{l=1}^M\left(L_{l,o}\operatorname{sign}(L_{o,k})\sqrt{\frac{L_{o,k}}{L_{k,o}}}+L_{l,k}\sqrt{L_{k,o}L_{o,k}}\right)^2\\
    &+\left(L_{l,k}\operatorname{sign}(L_{k,o})\sqrt{\frac{L_{k,o}}{L_{o,k}}}+L_{l,o}\sqrt{L_{k,o}L_{o,k}}\right)^2\Bigg)^{1/2}.
\end{align*}
Applying the approximations discussed in section~\ref{sec:coll_effects} to this equation, this expression reduces to the following equation,
 \begin{align*}
      \xi(o,k)&\approx \Lambda(o,k)\left( \frac{L_{o,k}}{L_{k,o}} \sum_{l=1}^M \left(L_{l,o}\right)^2 + \frac{L_{k,o}}{L_{o,k}} \sum_{l=1}^M \left(L_{l,k}\right)^2 \right)^{1/2}.
 \end{align*}
Based on this expression, we discuss certain limiting cases which explain the performance of the predictor. Assume that $L_{o,k}$ is very small keeping $L_{k,o}$ constant and much larger than $L_{o,k}$. In this case, the expression is dominated by $\xi(o,k)\approx |L_{k,o}|\left(\sum_{l=1}^M (L_{l,k})^2\right)^{1/2}$ which is predicted well by $L_{k,o}$. Performing the same approximation for small values of $L_{k,o}$ keeping $L_{o,k}$ constant and large, the expression is well predicted by $L_{o,k}$. For this reason, we need to keep both values in order to predict the overall collective effects. On the other hand, one can easily check that the approximation is equally valid if both LODFs are of the same order. 

Importantly, considering an arbitrary $\ell^p$-norm instead, the conclusions differ only slightly. An $\ell^p$ norm $\lVert \bm{x}\rVert_p$ of an arbitrary vector $\bm{x}\in\mathbb{R}^N$ is defined as 
$$\lVert \bm{x}\rVert_p = \left(\sum_{i=1}^N |x_i|^p\right)^{1/p}.$$ 
For the predictor, we then have
\begin{align*}
       &\xi_p(o,k) = \Lambda(o,k)\cdot\\  &\mathcal{L}(o,k)\Bigg(\sum_{l=1}^M\left|L_{l,o}\operatorname{sign}(L_{o,k})\sqrt{\frac{L_{o,k}}{L_{k,o}}}+L_{l,k}\sqrt{L_{k,o}L_{o,k}}\right|^p\\
    &+\left|L_{l,k}\operatorname{sign}(L_{k,o})\sqrt{\frac{L_{k,o}}{L_{o,k}}}+L_{l,o}\sqrt{L_{k,o}L_{o,k}}\right|^p\Bigg)^{1/p}.
\end{align*}
Therefore, the general expression to be considered does not change fundamentally when calculating the $\ell^p$ norm instead.

\section{Proof of proposition ~\ref{prop:lambdaproof}}\label{sec:proof}
\begin{proof}
The collectivity parameter defined by Eq.~\eqref{eq:collectivity_parameter} reads as 
\begin{align*}
    \xi(o,k) &= \mathcal{L}(o,k) \Big[\sum_{l=1}^M \left((L_{l,o}+L_{l,k}L_{k,o})L_{o,k}\right)^2+ \\
     &\left((L_{l,k}+L_{l,o}L_{o,k})L_{k,o}\right)^2\Big]^{1/2}.
\end{align*}
We will demonstrate that $\xi(o,k)$ is bounded from below by $\Lambda(o,k)$. First, since all addends in the sum are greater than zero, neglecting any or all of them will not increase the expression's value. We can thus choose the addends with $l=o$ and $l=k$. 

The expression then reads as
\begin{align*}
   \xi(o,k)= &\mathcal{L}(o,k) \Big[\sum_{l=1}^M \left((L_{l,o}+L_{l,k}L_{k,o})L_{o,k}\right)^2+ \\
     &\left((L_{l,k}+L_{l,o}L_{o,k})L_{k,o}\right)^2\Big]^{1/2}\\
    \geq\mathcal{L}(o,k)&\Big[\left((L_{o,o}+L_{o,k}L_{k,o})L_{o,k}\right)^2+\\ &\left((L_{k,k}+L_{k,o}L_{o,k})L_{k,o}\right)^2\Big]^{1/2}\stackrel{!}{\geq}\Lambda(o,k)\\
\end{align*}
Now we can make use of the fact that $L_{o,o}=L_{k,k}=-1$. In order to show that this expression is bounded from below by $\Lambda(o,k)$, we can square both sides of the inequality since all expressions considered here are positive. This yields
\begin{align*}
\xi(o,k)^2\geq& \mathcal{L}(o,k)^2\Big[\left((L_{o,k}L_{k,o}-1)L_{o,k}\right)^2+\\ &\left((L_{k,o}L_{o,k}-1)L_{k,o}\right)^2\Big]\\
     =&\mathcal{L}(o,k)^2 \frac{1}{\mathcal{L}(o,k)^2}\left(L_{o,k}^2+L_{k,o}^2\right)\\
     =& \left(L_{o,k}^2+L_{k,o}^2\right)\geq \Lambda(o,k)^2
\end{align*}
The last inequality follows from the following considerations 
\begin{align*}
    \left(L_{o,k}-L_{k,o}\right)^2+L_{o,k}L_{k,o}&> 0\\
    \Leftrightarrow  L_{o,k}^2+L_{k,o}^2-L_{o,k}L_{k,o}&> 0\\
    \Leftrightarrow  L_{o,k}^2+L_{k,o}^2&> L_{o,k}L_{k,o}=\Lambda(o,k)^2,
\end{align*}
which completes the proof. For $L_{o,k},L_{k,o}\neq 0$, the inequality is strict.
\end{proof}

\bibliography{lodf}
\end{document}